\documentclass[a4paper,11pt,final]{article}

\usepackage{graphicx,amsmath,amssymb,algorithm,algorithmic,textcomp,tikz-cd,float} 
\usepackage{epsfig}

\begin{document}

\title{\ \\ \LARGE\bf Fixation properties of multiple cooperator configurations on regular graphs}

\author{Hendrik Richter \\
HTWK Leipzig University of Applied Sciences \\ Faculty of
Electrical Engineering and Information Technology\\
        Postfach 301166, D--04251 Leipzig, Germany. \\ Email: 
hendrik.richter@htwk-leipzig.de }

\maketitle

\begin{abstract}
Whether or not cooperation is favored in  evolutionary games on graphs depends on the population structure and spatial properties of the interaction network. 
The population structure can be expressed as configurations. Such configurations extend scenarios with a single cooperator among defectors to any number of cooperators and any arrangement of cooperators and defectors on the network. For interaction networks modeled as regular graphs and for weak selection, the emergence of cooperation can be assessed by structure coefficients, which can be specified for each configuration and each regular graph. 
Thus, as a single cooperator can be interpreted as a lone mutant, the configuration--based structure coefficients also describe fixation properties of multiple mutants. 
We analyze the structure coefficients and particularly show that under certain conditions the coefficients strongly correlate to the average shortest path length between cooperators on the evolutionary graph. Thus,   
 for multiple cooperators fixation properties on regular evolutionary graphs can be linked to cooperator path lengths.

\end{abstract}

\section{Introduction}
A central issue in evolutionary game theory is describing conditions for the emergence of cooperation~\cite{broom13,nowak06}.
For games on evolutionary graphs with a single cooperator, the question of
whether or not cooperation is favored over defection is generally settled for weak selection. 
The answer is given for regular graphs by a single quantity, the structure coefficients~\cite{allen14,ohts06,tarnita09, taylor07}, while for graphs in general a quantity related to coalescence times can be used~\cite{allen17}. If there are multiple cooperators, the situation is more complicated~\cite{hinder14,pal10}. Recently, an approach has been proposed that analyses a given number of cooperators and defectors on the evolutionary graph as  configurations describing any possible arrangement of strategies over players ~\cite{chen13,chen16,rich17,rich18}. The approach also defines structure coefficients applicable for each configuration  \emph{and} for each network of interaction modeled by regular graphs~\cite{chen16}.

Game dynamics with multiple cooperators may be interesting for at least two reasons. A frequently studied question  is the appearance of a single cooperator by mutation that may or may not fixate in a frequency--dependent strategy updating process~\cite{broom11,chen13,ohts06,mcavoy16,patt15,tarnita09, taylor07}.  If, starting from a single cooperator configuration, fixation of cooperation succeeds, then the evolutionary process may undergo a transition through configurations with multiple cooperators.   Thus, studying multiple cooperator configurations can be seen as an attempt to describe how the transition process creates evolutionary trajectories.  Moreover, in social networks, it may be interesting to actively design conditions positive for cooperation. Such a design could involve the strategic placement of more than one cooperator to enhance the likelihood that cooperation prevails.

In this paper results are presented suggesting that for regular graphs fixation properties for any number of cooperators are strongly correlated to the cooperator path length, which is the average shortest path length between cooperators on the evolutionary graph.  In other words, we can link a property of the evolutionary game (emergence of cooperation) to a property of the graph (a certain path length). Moreover, it can be observed that particularly large values of the structure coefficient can be found for the cooperators clustering on cycles of the interaction graph. This result implies that clustering is certainly helpful for cooperation to succeed and suggests  the conjecture that graphs with abundant and overlapping cycles of length up to the number of cooperators produce more favorable  structure coefficients than graphs with few or separate cycles.   
We also discuss that these results are most important for small and intermediate numbers of players as the variance of structure coefficients over configurations ceases for the number of players getting large. 

The paper is structured as follows. In Sec.  \ref{sec:methods} the methodological framework of configuration--based structure coefficients is briefly recalled, see~\cite{chen16,rich18} for a more detailed description.  The main results are given in Sec. \ref{sec:results}, where we analyze  and compare evolutionary graphs covering a wide range of graph--theoretical properties.  It is shown how structure coefficients are distributed over configurations with the same number of cooperators and how structure coefficients correlate to cooperator path lengths. These relationships between fixation properties and cooperator path lengths relate to previous results 
for evolutionary games on lattice grids that show clusters of cooperators have better chances to survive~\cite{hau01,hau04,lang08,page00}. 
The findings are supported by results for networks with a varying number of players and coplayers. 
Finally, the results are discussed and conclusions are drawn about finding or designing best conditions to favor cooperation in evolutionary games on regular graphs.  

\section{Methods} \label{sec:methods}

We consider coevolutionary games with $N$ players $\mathcal{I}=\left(\mathcal{I}_1,\mathcal{I}_2,\ldots,\mathcal{I}_N\right)$.  Each pairwise interaction between 2 of these players, $\mathcal{I}_i$ and $\mathcal{I}_j$, who  thus are mutual coplayers, provides them with a payoff depending on which of 2 strategies, $C_i$ or $D_i$ and $C_j$ or $D_j$, each player uses. Each player has $k$ coplayers with $2 \leq k \leq N-1$. A (possibly varying) network structure defines which player  interacts with whom.   Such games are specified by (i.) the payoff matrix, (ii.) the interaction network, and (iii.) the configuration of the game~\cite{chen16,rich17}. 
The $2 \times 2$ payoff matrix is      
\begin{equation} 
\bordermatrix{~ & C_j & D_j \cr
                  C_i & a & b \cr
                  D_i & c & d \cr}. \label{eq:payoff}
\end{equation}
The interaction network is represented by an evolutionary graph. Each player $\mathcal{I}_i$ belongs to a vertex $i$ and an edge connecting vertex $i$ and vertex $j$ shows that the players   $\mathcal{I}_i$ and  $\mathcal{I}_j$ are mutual coplayers~\cite{allen17,sha12}. Such an interaction  graph  is equivalent to the adjacency matrix $A_I \in [0,1]^{N \times N}$. Set  the element  $a_{ij}$ of $A_I$ to  $a_{ij}=1$ for an edge between vertex 
$i$ and vertex $j$, and let  $a_{ij}=0$ indicate that the players   $\mathcal{I}_i$ and  $\mathcal{I}_j$ are not coplayers. We consider coevolutionary games with no self--play where each player  $\mathcal{I}_i$ has the same number $k$ of coplayers. Thus, the interaction graph is a simple $k$--regular graph.
Finally, a configuration $\pi=(\pi_1\pi_2\ldots\pi_N)$ specifies the strategy $\pi_i \in \{C_i,D_i\}$ that each player $\mathcal{I}_i$, ($i=1,2,\ldots,N$), uses. With 2 strategies ($C_i$ and $D_i$) there are $2^N$ configurations.  These configurations enumerate all possible arrangements of cooperators and defectors among the players. Additionally, the configurations  describe  any outcome of a player changing its strategy in a strategy updating process, for instance death--birth (DB) or birth--death (BD) updating~\cite{allen14,patt15,sha12}. 
It is convenient to binary code the strategies $\{C_i,D_i\} \rightarrow \{1,0\}$, thus having a binary string to specify the strategies of all players~\cite{chen16,rich17}. Consider the example of $N=4$ players. There are $2^4=16$ configurations. For instance, the configuration 
$\pi=(\pi_1\pi_2\pi_3\pi_4)=(0110)$   shows that players $\mathcal{I}_2$ and $\mathcal{I}_3$ cooperate, while    $\mathcal{I}_1$ and $\mathcal{I}_4$ defect.  As cooperation is coded as $C_i=1$, the Hamming weight of the string $\pi$ (the count of 1--elements in $\pi$) gives the number of cooperators $c(\pi)$ of a configuration $\pi$. Thus,  $\pi=(0110)$ has $c(0110)=2$ cooperators.  For $1 \leq c(\pi) \leq N-1$, there is more than one configuration $\pi$ for the same number of cooperators $c(\pi)$. For instance, there are $N$ configurations for a single cooperator ($c(\pi)=1$).  
The number $\#_{c(\pi)}$ of configurations with the same number of cooperators $c(\pi)$ can be calculated by: \begin{equation} \#_{c(\pi)}=\left(\begin{array}{c} N \\ c(\pi) \end{array} \right) \label{eq:config_number} \end{equation} with $1 \leq c(\pi) \leq N-1$ and $\sum_{c(\pi)=1}^{N-1} \#_{c(\pi)} =2^N-2$.

Recently, it was shown by Chen et al.~\cite{chen16} that  for  $2 \times 2$ games with $N$ players, payoff matrix (\ref{eq:payoff}), any configuration $\pi$ of cooperators and defectors and for any interaction network modeled by a simple, connected, $k$--regular graph,
in the case of weak selection strategy $C_i$ is favored over $D_i$ if \begin{equation} \sigma(\pi) (a-d)> (c-b). \label{eq:cond} \end{equation}
We call the quantity $\sigma(\pi)$ the structure coefficient of the configuration $\pi$ implying that it may have different values for different arrangements of cooperators and defectors decribed by $\pi$. The  condition (\ref{eq:cond}) generalizes the same condition with $\sigma(\pi)=\sigma=\frac{(k+1)N-4k}{(k-1)N}$,
which applies to a single cooperator, and is independent of where on the network the single cooperator is initally located~\cite{lehmann07,nowak10,tarnita09,taylor07}. Furthermore,
is was shown that for weak selection, the matrix $A_I$ describing the replacement structure,  and DB as well as BD updating, the structure coefficient $\sigma(\pi)$ can be calculated with time complexity  $\mathcal{O}(k^2N)$. In particular, for DB updating we have  \begin{equation} 
\sigma(\pi)=\frac{N\left(1+1/k \right) \overline{\omega^1} \cdot \overline{\omega^0}-2\overline{\omega^{10}}-\overline{\omega^1 \omega^0} }{N\left(1-1/k \right) \overline{\omega^1} \cdot \overline{\omega^0}+\overline{\omega^1 \omega^0}},  \label{eq:sigma}
\end{equation} with 4 local frequencies:  $\overline{\omega^1}$, $\overline{\omega^0}$, $\overline{\omega^{10}}$ and $\overline{\omega^1 \omega^0}$, see~\cite{chen16}. For these local frequencies, the following probabilistic interpretation has been suggested. Suppose on a given interaction network a random walk is carried out with the starting 
vertex chosen uniformly--at--random. The local frequency $\overline{\omega^1}$ (or $\overline{\omega^0}=1-\overline{\omega^1}$) is the probability that for a configuration $\pi$ the player at the first step of the walk is a cooperator (or defector). The local frequency  $\overline{\omega^{10}}$ is the probability that for a walk with 2 steps the player at the first step is  a cooperator and at the second step it is a defector. Finally, if we carry out 2 random walks independent of each other, the local frequency   $\overline{\omega^1 \omega^0}$ is the probability that the player at the first step on the first walk is a cooperator, but a defector at the first step on the second walk.

To define the cooperator path length $l_c$
we call $d_{ij}(\mathcal{I}_i,\mathcal{I}_j)$ the shortest path on the evolutionary graph  between vertex $i$ (= player $\mathcal{I}_i$) and vertex $j$ (= player $\mathcal{I}_j$) given that both players  are cooperators: $\pi_i=\pi_j=1$. The quantity $d_{ij}(\mathcal{I}_i,\mathcal{I}_j)$ can be calculated by variants of Dijkstra's algorithm with time complexity $\mathcal{O}(kN/2+N\log{(N)})$. If there are more than 2 cooperators, $c(\pi)>2$, there 
are $c(\pi)(c(\pi)-1))/2>1$ pairs  with individual path lengths  $d_{ij}(\mathcal{I}_i,\mathcal{I}_j)$. The  cooperator  path length $l_c$ is  the mean over all pairs: \begin{equation} l_c=\frac{2}{c(\pi)(c(\pi)-1)} \sum_{i,j \atop i\neq j} d_{i,j}(\mathcal{I}_i,\mathcal{I}_j). \label{eq:av_path} \end{equation}


\begin{figure}[tb]
\includegraphics[trim = 10mm 110mm 0mm 80mm,clip,width=17.25cm, height=9.5cm]{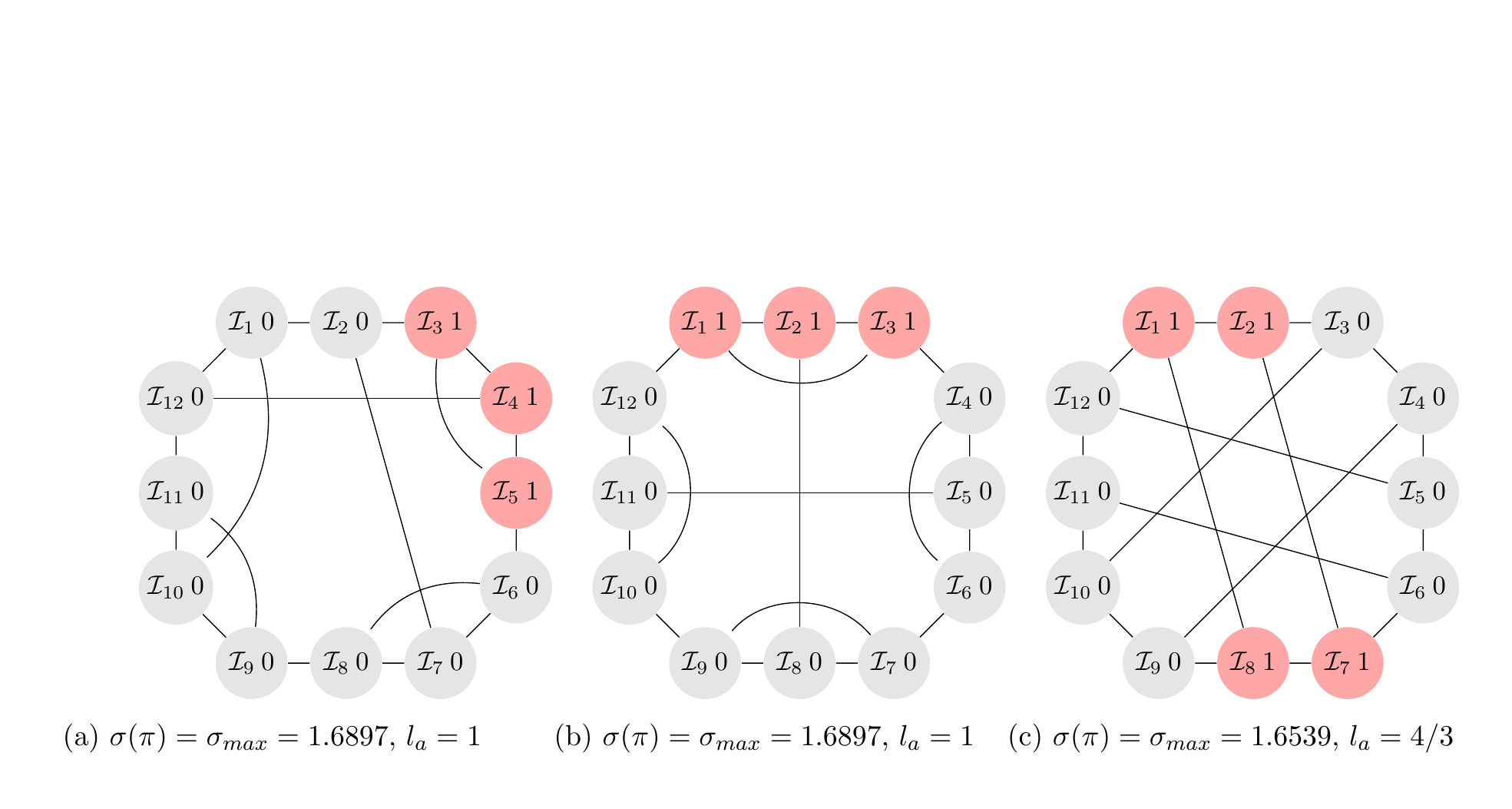}

\caption{\small{Configurations $\pi$, structure coefficients $\sigma(\pi)$ and cooperator path lengths $l_c$ for: (a) the Frucht~\cite{chen16,fru49,mcavoy15}, (b) the truncated tetrahedral~\cite{read98} and (c) the Franklin graph~\cite{frank34}. The truncated tetrahedral and the Franklin graph are vertex--transitive, while the Frucht graph is not, but possesses  no non--trivial automorphism. All these graphs are Hamiltonian.  Configurations with the maximal structure coefficient $\sigma_{max}$ are shown for: (a),(b) $c(\pi)=3$ cooperators; (c) $c(\pi)=4$ cooperators. }}

\label{fig:loc_frucht0}
\end{figure}

\section{Results} \label{sec:results}
\begin{figure}[tb]
\includegraphics[trim = 10mm 110mm 0mm 80mm,clip,width=17.25cm, height=9.5cm]{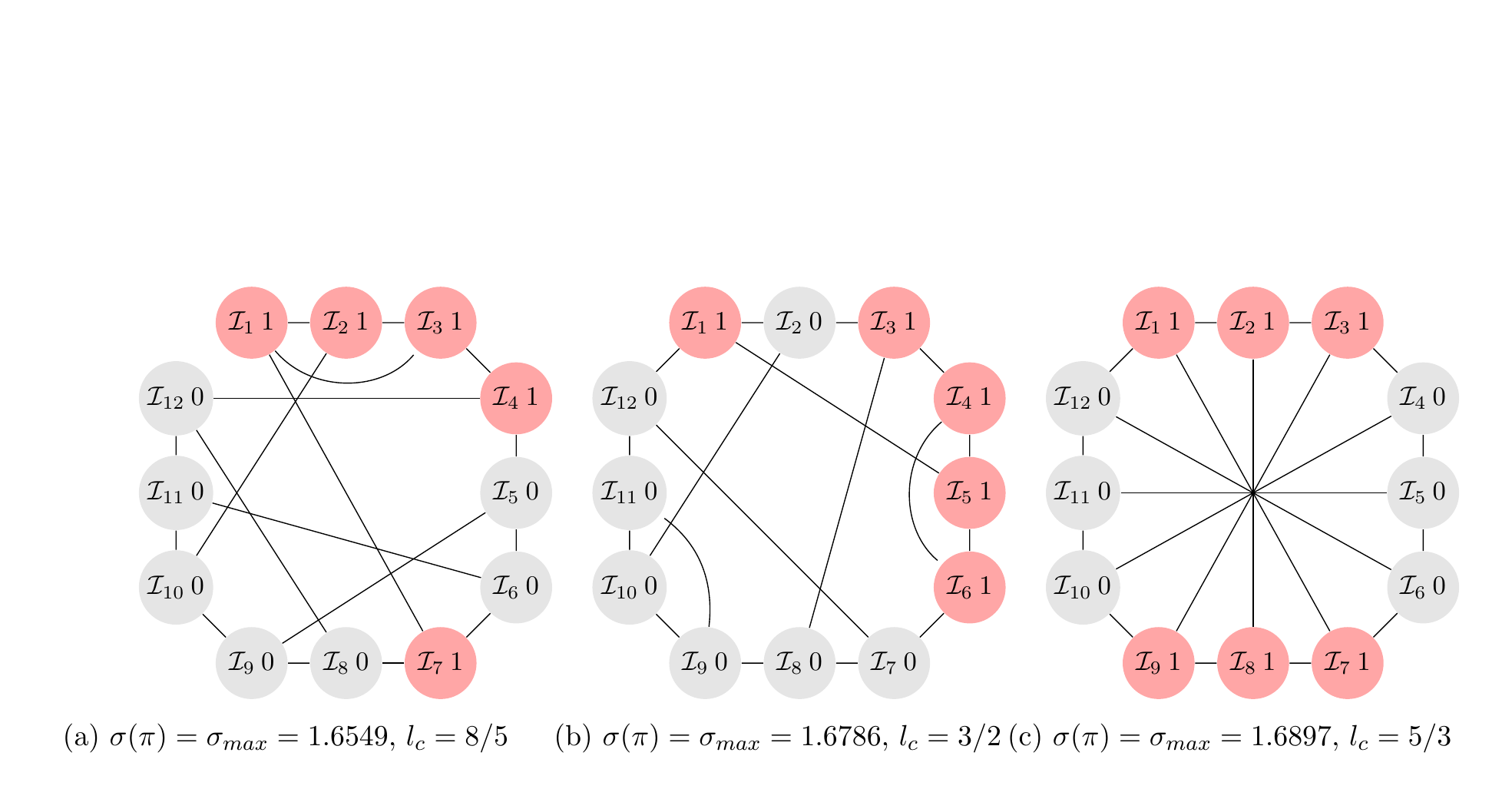}

\caption{\small{Configurations $\pi$, structure coefficients $\sigma(\pi)$ and cooperator path lengths $l_c$ for: (a) the Tietze, (b) the D\"urer and (c) the M\"obius ladder graph,~\cite{bond08,read98}. The M\"obius ladder graph is vertex--transitive, while the Tietze and the D\"urer graph are not, but both possess non--trivial automorphisms. The D\"urer graph is Hamiltonian, while the Tietze graph is not.  Configurations with the maximal structure coefficient $\sigma_{max}$ are shown for:  (a),(b) $c(\pi)=5$  cooperators; (c) $c(\pi)=6$ cooperators. }}

\label{fig:loc_tietze_duer}
\end{figure}

Studying  relationships between configurations $\pi$, structure coefficients $\sigma(\pi)$ and cooperator path lengths $l_c$ starts with 6 different interaction networks
with $N=12$ players and $k=3$ coplayers that represent a variety of graph properties, see Tab. \ref{tab:graph} and Figs. \ref{fig:loc_frucht0}--\ref{fig:loc_tietze_duer}, and also the Appendix 1 with Figs. \ref{fig:loc_fre}--\ref{fig:loc_4coop}. We analyze the following graphs:
Frucht~\cite{chen16,fru49,mcavoy15},  truncated tetrahedral~\cite{read98}, Franklin~\cite{frank34},  Tietze~\cite{bond08,mcavoy15}, D\"urer and M\"obius ladder~\cite{read98}.

 \begin{table}[H]
\caption{\small{Analysis of the structure coefficients $\sigma(\pi)$ and the cooperator path lengths $l_c$ for varying numbers of cooperators $2 \leq c(\pi) \leq 6$ and different graphs of size $N=1$2 and degree $k=3$. The table shows the multiplicity of the structure coefficient $\#_\sigma$, the maximal and minimal structure coefficient, $\sigma_{max}$ and $\sigma_{min}$, the multiplicity of the maximal and minimal structure coefficient, $\#_{\sigma_{max}}$ and  $\#_{\sigma_{min}}$  and the multiplicity of the cooperator path length, $\#_{l_{c\: min}}$ and  $\#_{l_{c\: max}}$. Note that the number of configurations with the same number of cooperators are
$\#_{c(\pi)}=\left (\#_2, \#_3, \#_4, \#_5, \#_6 \right)=\left (66,220,495,792,924 \right)$, which is calculated by the binomials $\#_{c(\pi)}=\frac{12!}{c(\pi)! (12-c(\pi))!}$, see Eq. (\ref{eq:config_number}). The truncated tetrahedral graph, the Franklin graphs and the M\"obius ladder graph are vertex--transitive. The multiplicities  $\#_{\sigma_{max}}$, $\#_{\sigma_{min}}$, $\#_{l_{c\: min}}$ and  $\#_{l_{c\: max}}$ are underlined if they match, which means $\#_{\sigma_{max}}=\#_{l_{c\: min}}$ and $\#_{\sigma_{min}}=\#_{l_{c\: max}}$ and implies the largest value of the structure coefficient $\sigma(\pi)$ is uniquely determined by the smallest value of the cooperator path length $l_c$, and vice versa. Furthermore, the results for the structure coefficients $\sigma(\pi)$ are symmetric with respect to the number of cooperators, compare to Fig. \ref{fig:viol_frucht_frank} (a),(b). This means the values for $\#_\sigma$, $\sigma_{max}$, $\sigma_{min}$, $\#_{\sigma_{max}}$  and  $\#_{\sigma_{min}}$ are the same for $c(\pi)=\{5,7\}$,  $c(\pi)=\{4,8\}$,  $c(\pi)=\{3,9\}$ and so on.  For the vertex--transitive graphs this symmetry also applies  for  $\#_{l_{c\: min}}$ and $\#_{l_{c\: max}}$.    }} \label{tab:graph}

\begin{tabular}{|c|@{}c@{}|}\hline
Graph & \begin{tabular}{p{1.5cm}p{1.5cm}|p{1cm}|p{1cm}p{1cm}p{1cm}p{1cm}p{1cm}p{1cm}p{1cm}} Triangles & Squares & $c(\pi)$ & $\#_\sigma$ & $\sigma_{max}$ & $\sigma_{min}$ & $\#_{\sigma_{max}}$ &  $\#_{\sigma_{min}}$ & $\#_{l_{c\: min}}$ & $\#_{l_{c\: max}}$\\   \end{tabular}
\tabularnewline\hline
\begin{tabular}{c}
Frucht \\
Fig. \ref{fig:loc_frucht0}(a)
\end{tabular}
&
\begin{tabular}{p{1.5cm}p{1.5cm}|p{1cm}|p{1cm}p{1cm}p{1cm}p{1cm}p{1cm}p{1cm}p{1cm}}
  &  & 2 & 5& 1.5846 & 1.4546 & 9 & 23 &18 & 2\\
  &  & 3 & 14& 1.6897 & 1.4000 & \underline{3} & 1 & \underline{3} & 2\\
 3 & 1& 4 & 20 & 1.7059 & 1.4231 & 2 & 23 & 10 & 4\\
  && 5& 24 & 1.7568 & 1.4210 & \underline{1} & 19 & \underline{1} & 5\\
&& 6 & 25 & 1.7130 & 1.4202 & 6 & 24& 3 & 3\\
\end{tabular}
\tabularnewline\hline
\begin{tabular}{c}
Truncated \\
tetrahedral \\
Fig. \ref{fig:loc_frucht0}(b)
\end{tabular}
&
\begin{tabular}{p{1.5cm}p{1.5cm}|p{1cm}|p{1cm}p{1cm}p{1cm}p{1cm}p{1cm}p{1cm}p{1cm}}
  &  & 2 & 4& 1.5846 & 1.4546 & 12 & \underline{24} & 18 & \underline{24}\\
  &  & 3 & 10& 1.6897 & 1.4000 & \underline{4} & \underline{4} & \underline{4} & \underline{4}\\
4  &0 & 4 & 12 & 1.6796 & 1.4231 & \underline{12} & 30 & \underline{12}& 24\\
  && 5 & 15 & 1.6549 & 1.4210 & 36 & 24 & 12 & 12\\
&& 6 & 18 & 1.6897 & 1.4000 & 6 & \underline{6} & 38 & \underline{6}\\
\end{tabular}
\tabularnewline\hline
\begin{tabular}{c}
Franklin \\
Fig. \ref{fig:loc_frucht0}(c)
\end{tabular}
&
\begin{tabular}{p{1.5cm}p{1.5cm}|p{1cm}|p{1cm}p{1cm}p{1cm}p{1cm}p{1cm}p{1cm}p{1cm}}
 &  & 2 & 4& 1.5455 & 1.4546 &  \underline{18} &  \underline{18} &  \underline{18} &  \underline{18}\\
  &  & 3 & 8& 1.5909 & 1.4270 & 12 & 24 & 36 & 36\\
  0&3 & 4 & 13 & 1.6539 & 1.4231 &  \underline{3} & 15 &  \underline{3} & 3\\
&& 5 & 17 & 1.6316 & 1.4210 &  \underline{12} &  \underline{12} &  \underline{12} &  \underline{12}\\
&& 6 & 18 & 1.6379 & 1.4000 &  \underline{6} & 8 &  \underline{6} & 6\\

\end{tabular}
\tabularnewline\hline
\begin{tabular}{c}
Tietze \\
Fig. \ref{fig:loc_tietze_duer}(a)
\end{tabular}
&
\begin{tabular}{p{1.5cm}p{1.5cm}|p{1cm}|p{1cm}p{1cm}p{1cm}p{1cm}p{1cm}p{1cm}p{1cm}}
 &  & 2 & 4& 1.5846 & 1.4546 & 3 & \underline{15} & 18 & \underline{15}\\
  &  & 3 & 11& 1.6897 & 1.4000 &  \underline{1} &  \underline{1} &  \underline{1} &  \underline{1}\\
  1&0 & 4 & 14 & 1.6796 & 1.4231 &  \underline{3} & 12 &  \underline{3} & 3\\
&& 5 & 20 & 1.6549 & 1.4210 & 3 & 12 & 6 & 18\\
&& 6 & 18 & 1.6154 & 1.4202 & 26 & 24 & 13 & 14\\

\end{tabular}
\tabularnewline\hline
\begin{tabular}{c}
D\"urer \\
Fig. \ref{fig:loc_tietze_duer}(b)
\end{tabular}
&
\begin{tabular}{p{1.5cm}p{1.5cm}|p{1cm}|p{1cm}p{1cm}p{1cm}p{1cm}p{1cm}p{1cm}p{1cm}}
 &  & 2 & 4& 1.5846 & 1.4546 & 6 & 18 & 18 & 3\\
  &  & 3 & 10& 1.6897 & 1.4270 &  \underline{2} & 12 &  \underline{2} & 42\\
  2&0 & 4 & 16 & 1.6796 & 1.4231 &  \underline{6} & 24 &  \underline{6} & 3\\
&& 5 & 20 & 1.6786 & 1.4210 & 6 & 18 & 12 & 36\\
&& 6 & 19 & 1.6897 & 1.4310 & 6 & 36 & 8 & 1\\

\end{tabular}
\tabularnewline\hline
\begin{tabular}{c}
M\"obius  \\ ladder\\
Fig. \ref{fig:loc_tietze_duer}(c)
\end{tabular}
&
\begin{tabular}{p{1.5cm}p{1.5cm}|p{1cm}|p{1cm}p{1cm}p{1cm}p{1cm}p{1cm}p{1cm}p{1cm}}
 &  & 2 & 4& 1.5455 & 1.4546 &  \underline{18} &  \underline{24} &  \underline{18} &  \underline{24}\\
  &  & 3 & 9& 1.5909 & 1.4000 & 24 &  \underline{4} & 36 &  \underline{4}\\
  0&6 & 4 & 17 & 1.6539 & 1.4340 &  \underline{6} &  \underline{48} &  \underline{6} &  \underline{48}\\
&& 5 & 20 & 1.6549 & 1.4310 &  \underline{24} & 60 &  \underline{24} & 24\\
&& 6 & 20 & 1.6897 & 1.4000 &  \underline{6} & 6 &  \underline{6} & 4\\

\end{tabular}
\tabularnewline\hline

\end{tabular}
\end{table}

 Of these graphs, 3 are just regular, but differ by having   no non--trivial symmetry (Frucht), possessing non--trivial automorphisms and being Hamiltonian (D\"urer) or not (Tietze). The other 3 graphs 
are vertex--transitive, which is to say that some automorphism group acts transitively upon the graph's vertices. Informally speaking, vertex--transitive means the graphs look the same from any two vertices and thus have a higher degree of symmetry than graphs that are just regular. Additionally, the truncated tetrahedral, Tietze and D\"urer graph have  triangles (cycles of length 3), but are square free, while the Franklin and  M\"obius  ladder graphs have squares (cycles of length 4), but are triangle free.

For these graphs large values of the structure coefficients $\sigma(\pi)$ correspond to small values of the cooperator path length $l_c$, and vice versa,   suggesting the hypothesis that    between the structure coefficients $\sigma(\pi)$ and the cooperator path length $l_c$ there is an inversely proportional relationship.   For example, the Frucht graph with 3 cooperators has the maximal structure coefficient 
$\sigma_{max}=1.6897$ for 3 configurations which all share that the 3 cooperators are distanced by the minimal $l_c=1$ \emph{and}  belong to one of the 3 triangles of the Frucht graph, see Fig. \ref{fig:loc_frucht0}(a) which shows one of these configurations. The same applies for 3 cooperators and the truncated tetrahedral graph, Fig. \ref{fig:loc_frucht0}(b), while  with 4 cooperators and the Franklin graph, Fig. \ref{fig:loc_frucht0}(c), we get 3 configurations with $\sigma_{max}=1.6539$ and the minimal $l_c=4/3$.  
A more detailed analysis of configurations $\pi$, structure coefficients $\sigma(\pi)$ and cooperator path lengths $l_c$ for these interaction networks is given in the Appendix 1 with Figs. \ref{fig:loc_fre}--\ref{fig:loc_4coop}.
Particularly, it is shown that for 4 cooperators, the largest or smallest value of $l_c$ in itself does not guarantee the  smallest or largest value of $\sigma(\pi)$.
For instance, for the Franklin graph and 4 cooperators, $c(\pi)=4$,  not only the smallest value $l_c=4/3$, but also that the 4 cooperators form one of the squares, gives the largest structure coefficient $\sigma_{max}=1.6539$, Fig. \ref{fig:loc_frucht0}(c), while for the truncated tetrahedral graph, the minimal $\sigma_{min}=1.4231$ is obtained for the second largest value $l_c=3$, Fig. \ref{fig:loc_4coop}(e).
 It can further be observed  that the maximal and minimal value of $\sigma(\pi)$ is rarely unique for a configuration $\pi$, even if the number of cooperators $c(\pi)$ is constant.


In Tab. \ref{tab:graph}, the largest and smallest $\sigma(\pi)$, denoted as $\sigma_{max}$ and $\sigma_{min}$, are given for each $2 \leq c(\pi) \leq 6$. 
 The table also shows the multiplicity of the structure coefficients $\#_\sigma$, and the multiplicity of the maximal and minimal structure coefficient, $\#_{\sigma_{max}}$ and $\#_{\sigma_{min}}$ as well as   the multiplicity of the maximal and minimal cooperator path length, $l_{a \: max}$ and $l_{a \: min}$.  
The multiplicity of the structure coefficient $\#_\sigma$ is the number of different values of $\sigma(\pi)$. For instance, for the Frucht graph and $c(\pi)=6$, $\#_\sigma=25$ means there are 25 different values of $\sigma(\pi)$ over the $\#_6=924$ configurations with 6 cooperators.  Similarly, the multiplicity of the maximal structure coefficient $\#_{\sigma_{max}}$ is the number of configurations that have $\sigma(\pi)=\sigma_{max}$. Thus,  for the Franklin graph and $c(\pi)=5$,  $\#_{\sigma_{max}}=12$ means 12 out of 792 configurations have the maximal structure coefficient $\sigma_{max}=1.6316$. For $\#_{\sigma_{min}}$, 
$\#_{l_{c\: min}}$ and $\#_{l_{c\: max}}$ this applies likewise.

Looking at the results in Tab. \ref{tab:graph},
we see some patterns over the 6 graphs considered, but also some differences. 
A first pattern is that the multiplicity of the structure coefficients $\#_\sigma$ does not scale to the number of configurations $\#_{c(\pi)}$ for each number of cooperators $c(\pi)$.  Clearly, for $c(\pi)=2$ and $\#_2=66$ configurations with 2 cooperators, we also get the smallest $\#_\sigma$ in all cases given in Tab. \ref{tab:graph}, but as $c(\pi)$ and  $\#_{c(\pi)}$ increases, the multiplicity of the structure coefficient  $\#_\sigma$ does not. The same property can also be found for the multiplicity of maximal and minimal $\sigma(\pi)$ and $l_c$. Another interesting result is that generally the multiplicity of the minimal $\sigma_{min}$ is larger than for the maximal $\sigma_{max}$,  that is, mostly we have $\#_{\sigma_{max}} \leq \#_{\sigma_{min}}$. A possible interpretation is that finding a configuration $\pi$ at random with the maximal structure coefficient $\sigma_{max}$ is less likely than finding a configuration with $\sigma_{min}$. However, this interpretation should be viewed with respect to the number of configurations with the same number of cooperator $\#_{c(\pi)}$. 
For instance, for the Frucht graph, we get for 2 cooperators ($c(\pi)=2$) the values $\#_{\sigma_{max}}=9$ and $\#_{\sigma_{min}}=23$. With $\#_2=66$ configurations with 2 cooperators, the likelihood to find at random a configuration with   $\sigma_{max}$ is $9/66=0.1364$, while for   $\sigma_{min}$ it is $23/66=0.3485$.  However, if we consider the same for 6 cooperators, we get $6/924=0.0065$ and $24/924=0.0260$ for $\sigma_{max}$ and  $\sigma_{min}$. In other words, looking at random for configurations with certain fixation properties might be  promising for a small number of cooperators, but surely it is not for a larger $c(\pi)$.

Comparing the graphs the most interesting result is that for the vertex--transitive graphs (truncated tetrahedral, Franklin and M\"obius ladder) the multiplicity of $\sigma_{max}$ and $\sigma_{min}$  matches the multiplicity of  $l_{a \: min}$ and $l_{a \: max}$  more often than for the other graphs, see Tab.  \ref{tab:graph}, where in the columns for $\#_{\sigma_{max}}$, $\#_{\sigma_{min}}$, $\#_{l_{c\: min}}$ and  $\#_{l_{c\: max}}$ the matching values are underlined. This means for the vertex--transitive graphs considered it is more likely that the largest value of the structure coefficient $\sigma(\pi)$ is uniquely determined by the smallest value of the cooperator path length $l_c$, and vice versa.   While for the Frucht, Tietze and D\"urer graph, there are 2, 4 and 2 matches, we get for the tetrahedral, Franklin and  M\"obius ladder graph 5, 6 and 7 matches. Thus, we can hypothesize   that for vertex--transitive graphs the correlation between structure coefficients and cooperator path lengths might be stronger than for graphs that are just regular. The next set of numerical results deals with this hypothesis and  generally with the proposed inverse proportional relationships between structure coefficients and cooperator path lengths. 

\newsavebox{\smlmat}
\savebox{\smlmat}{$\left(\begin{smallmatrix}a &b\\c & d \end{smallmatrix}\right)=\left(\begin{smallmatrix}4 &0\\5 & 1 \end{smallmatrix}\right)$}

\begin{figure}[htb]

\includegraphics[trim = 0mm 0mm 0mm 0mm,clip,width=8.25cm, height=5.9cm]{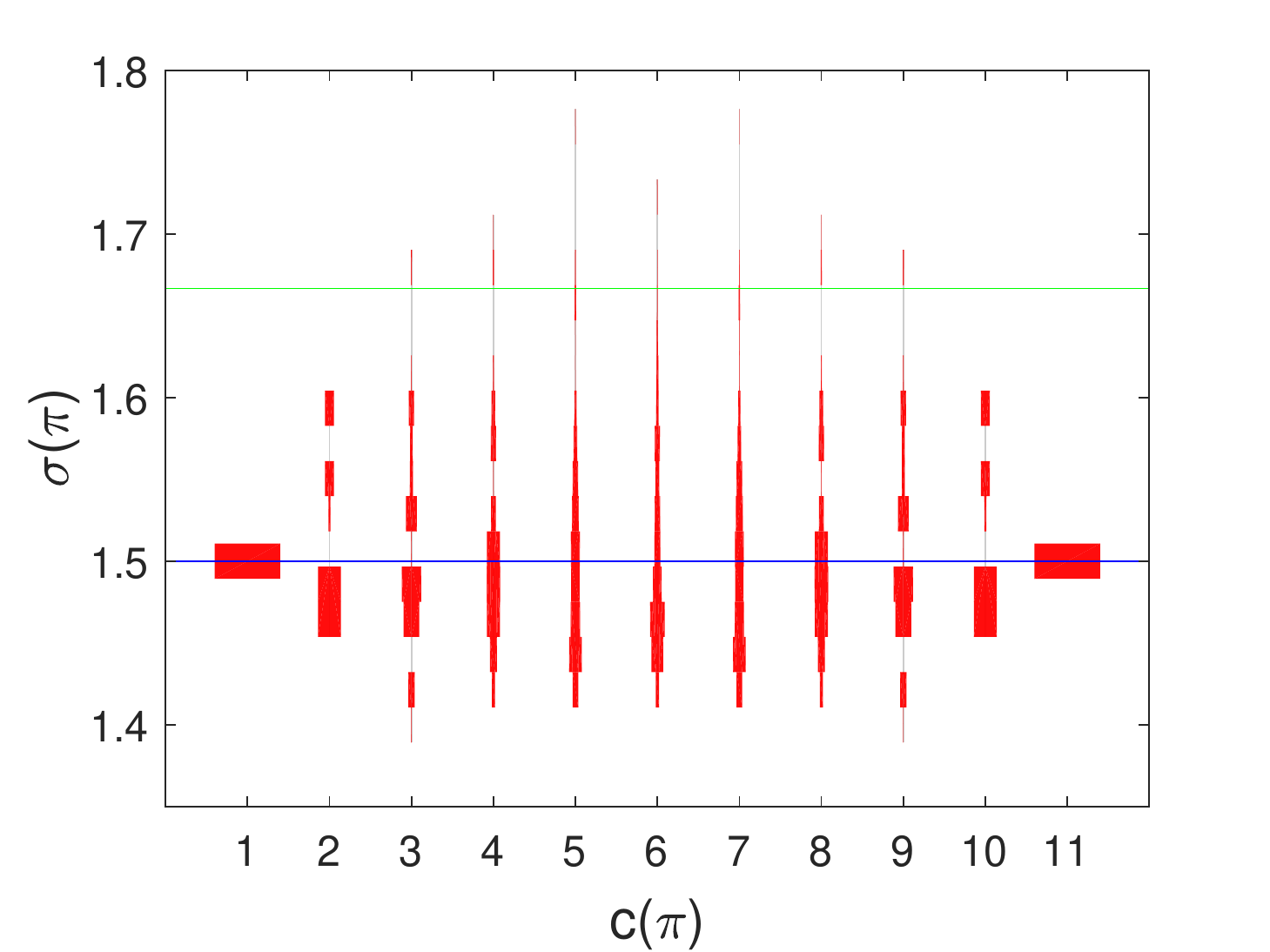}
\includegraphics[trim = 0mm 0mm 0mm 0mm,clip,width=8.25cm, height=5.9cm]{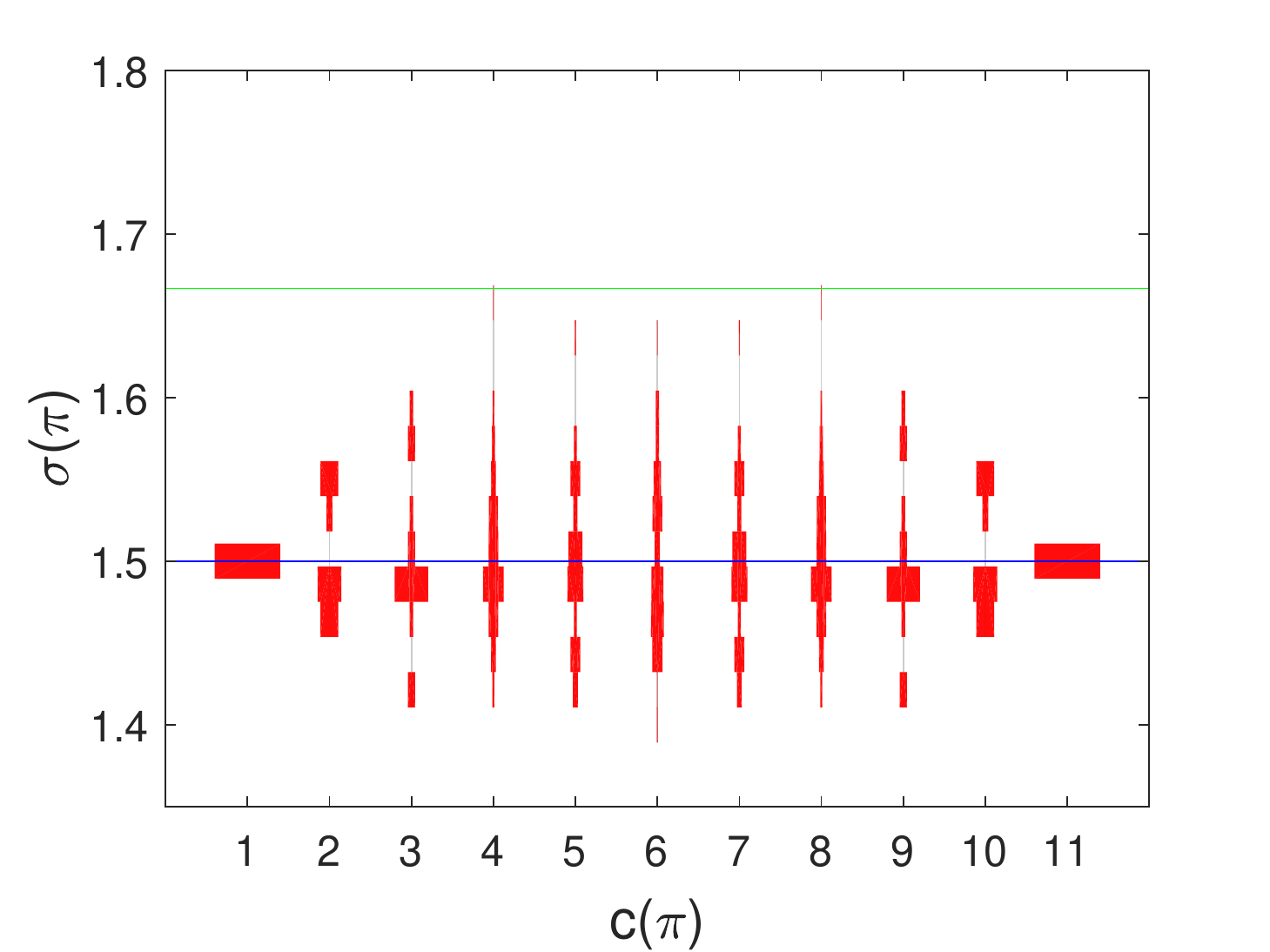}

\hspace{1cm} (a)  \hspace{7cm} (b) 

\includegraphics[trim = 0mm 0mm 0mm 0mm,clip,width=8.25cm, height=5.9cm]{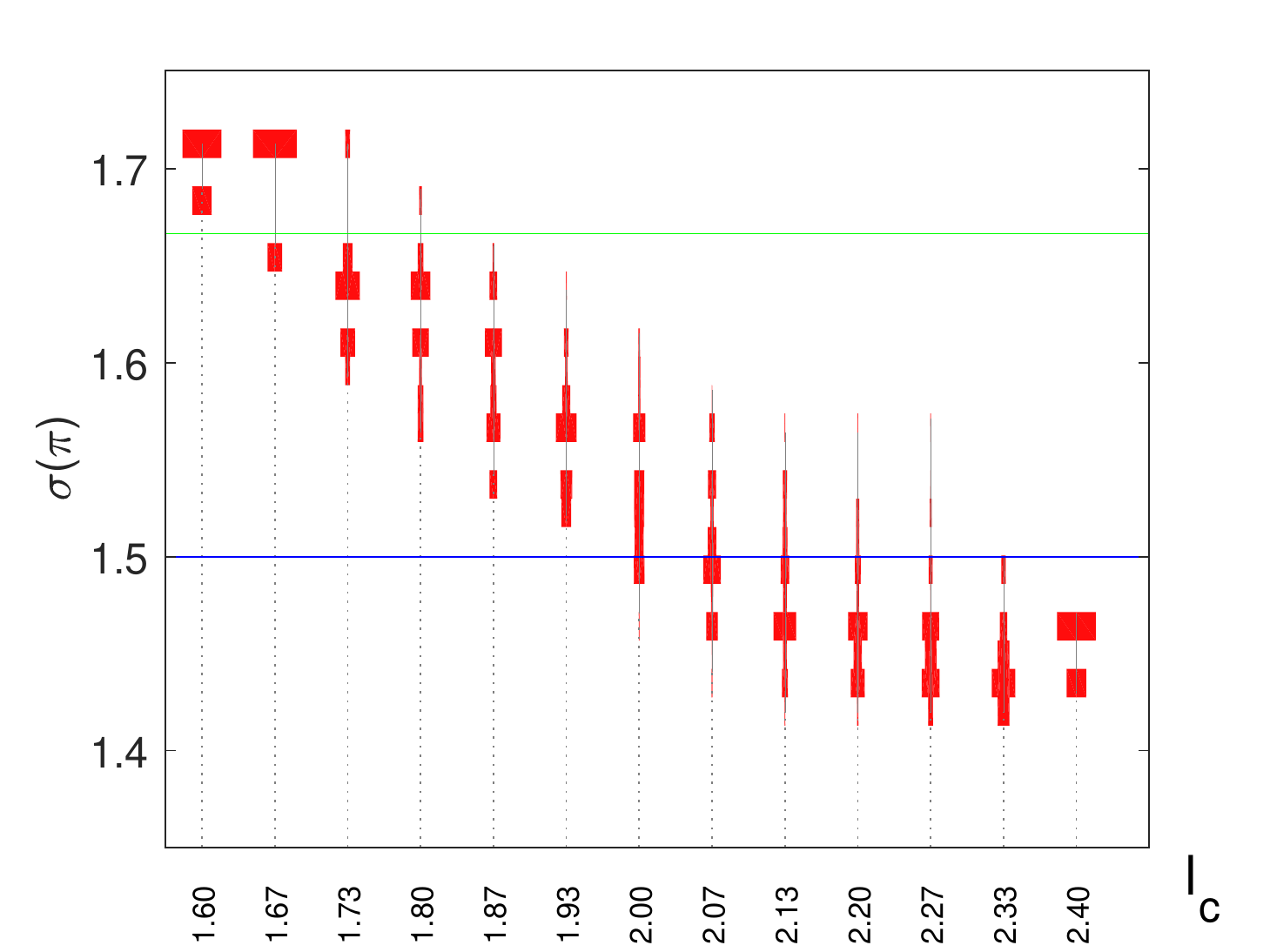}
\includegraphics[trim = 0mm 0mm 0mm 0mm,clip,width=8.25cm, height=5.9cm]{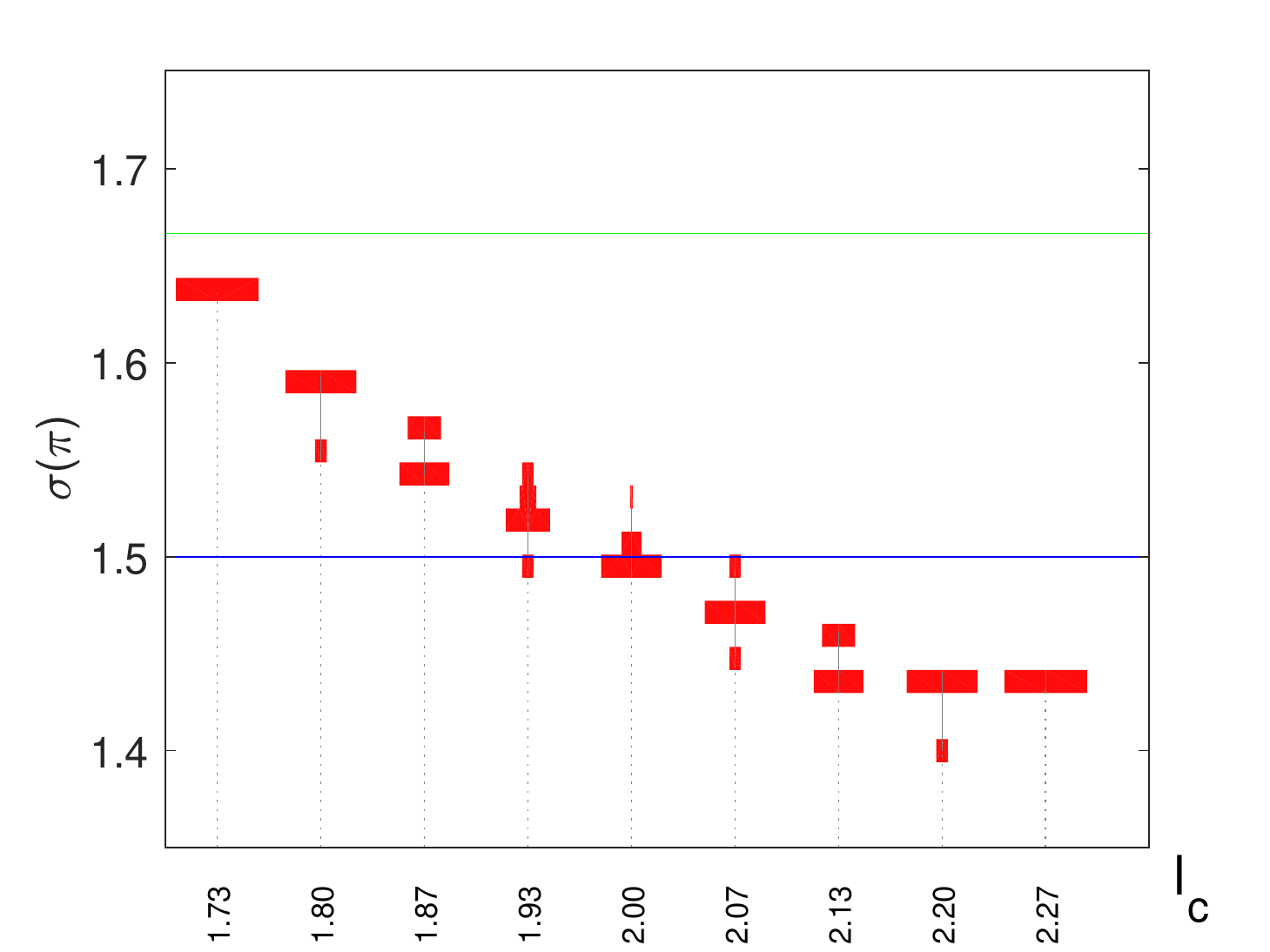}

\hspace{1cm} (c)  \hspace{7cm} (d) 

\vspace{2mm}

\caption{Distributions of the structure coefficients $\sigma(\pi)$ for the Frucht graph (a),(c) and the Franklin graph (b),(d) over the number of cooperators $c(\pi)$ (a),(b) and unique values of the cooperator path length $l_c$ for $c(\pi)=6$ (c),(d).   The blue lines intersecting the distributions show the value $\sigma=1.5$ which is obtained for a single cooperator with $N=12$, $k=3$ and $\sigma=\frac{(k+1)N-4k}{(k-1)N}$. The green lines indicate the condition $\sigma(\pi)>5/3$  to favor cooperation in a PD game with~\usebox{\smlmat}. }
\label{fig:viol_frucht_frank}
\end{figure}

\begin{figure}[tb]
\includegraphics[trim = 0mm 30mm 0mm 25mm,clip,width=17.25cm, height=5.9cm]{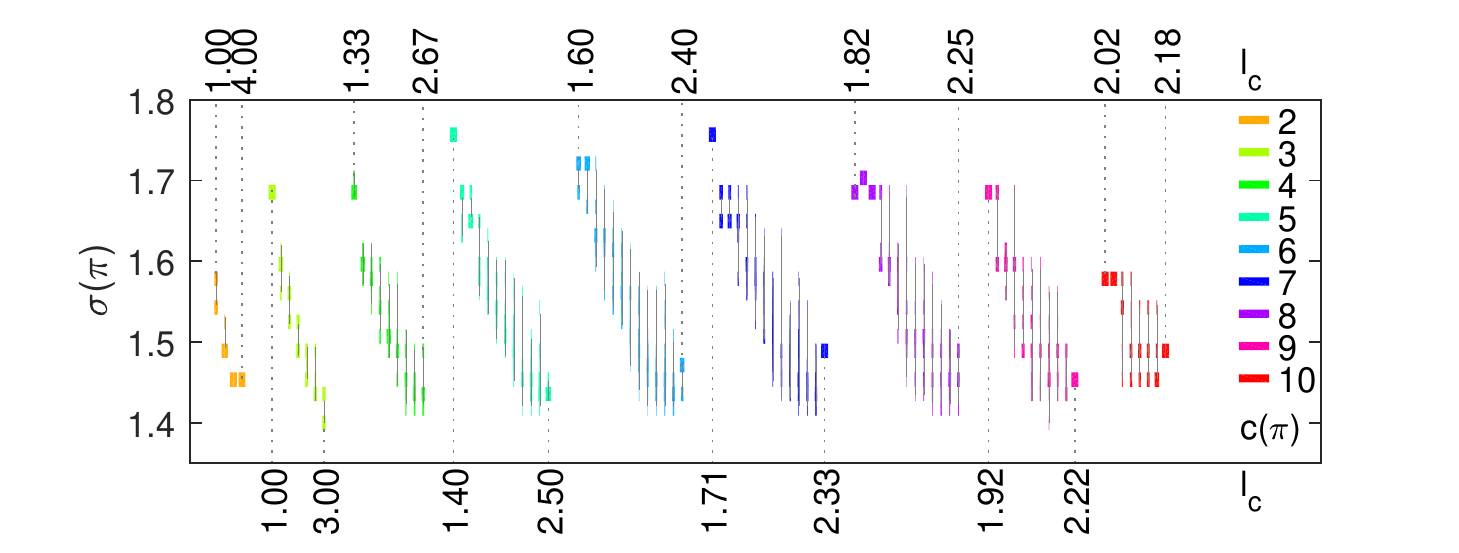}

\hspace{1cm} (a) Frucht graph, Fig. \ref{fig:loc_frucht0}(a)

\includegraphics[trim = 0mm 30mm 0mm 25mm,clip,width=17.25cm, height=5.9cm]{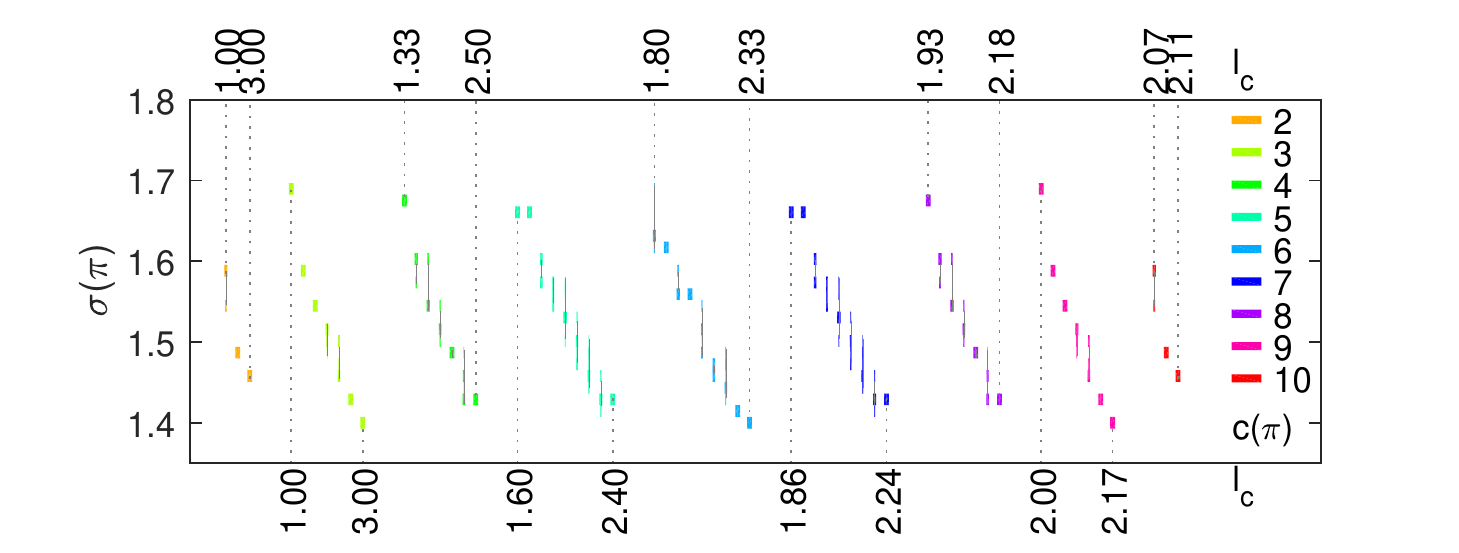}

\hspace{1cm} (b) Truncated tetrahedral graph, Fig. \ref{fig:loc_frucht0}(b)

\includegraphics[trim = 0mm 30mm 0mm 25mm,clip,width=17.25cm, height=5.9cm]{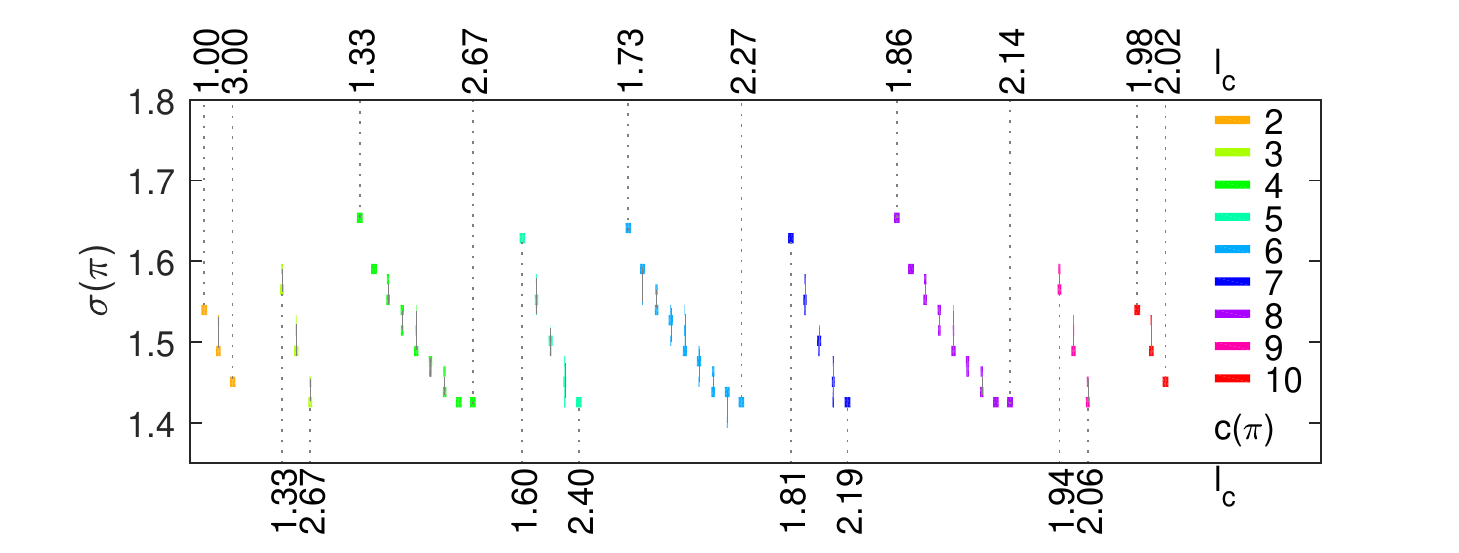}

\hspace{1cm} (c) Franklin graph, Fig. \ref{fig:loc_frucht0}(c)

\end{figure}

\begin{figure}[tb]
\includegraphics[trim = 0mm 30mm 0mm 25mm,clip,width=17.25cm, height=5.9cm]{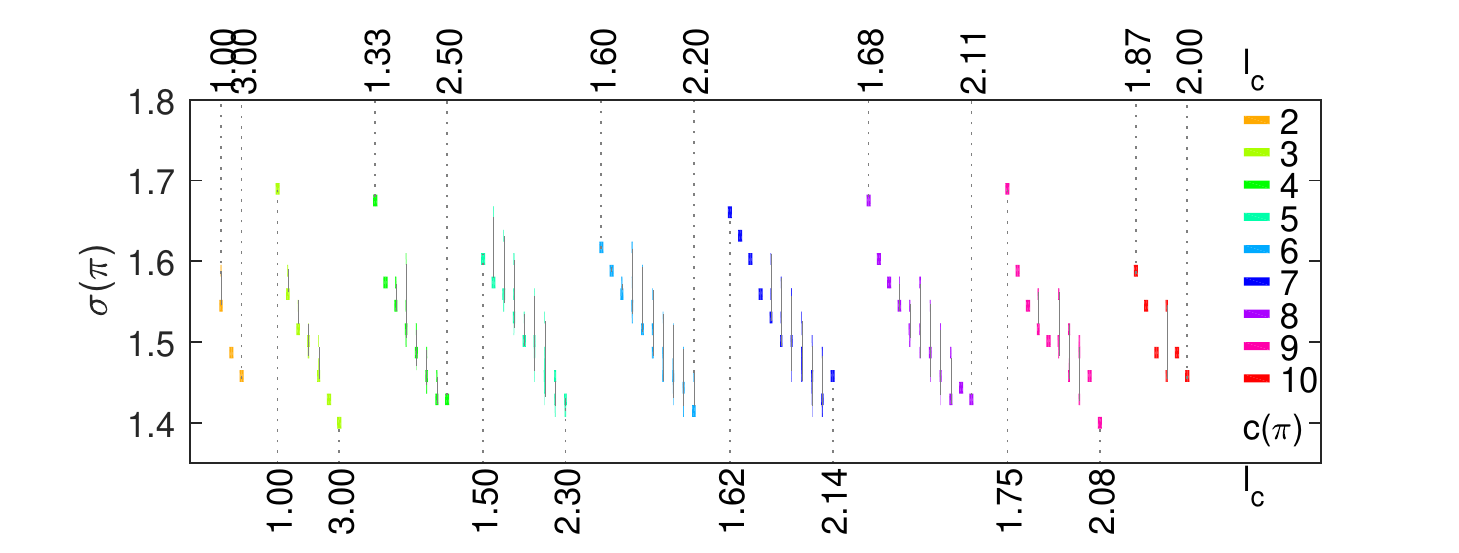}

\hspace{1cm} (d) Tietze graph, Fig. \ref{fig:loc_tietze_duer}(a)

\includegraphics[trim = 0mm 30mm 0mm 25mm,clip,width=17.25cm, height=5.9cm]{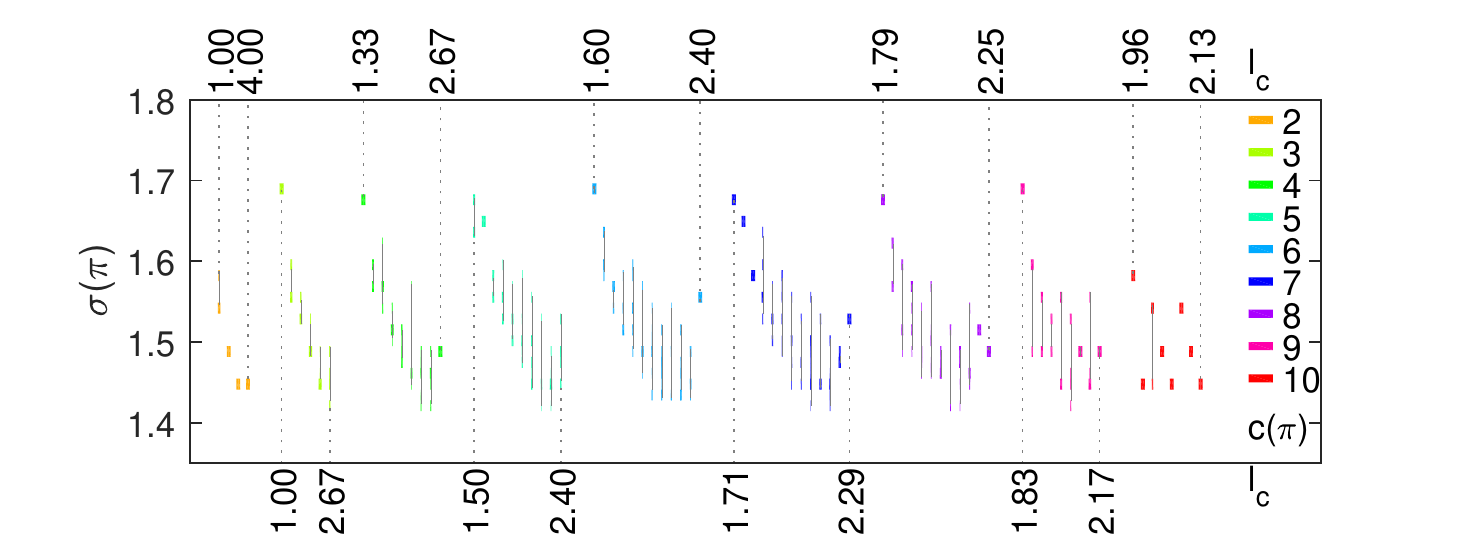}

\hspace{1cm} (e) D\"urer graph, Fig. \ref{fig:loc_tietze_duer}(b)

\includegraphics[trim = 0mm 30mm 0mm 25mm,clip,width=17.25cm, height=5.9cm]{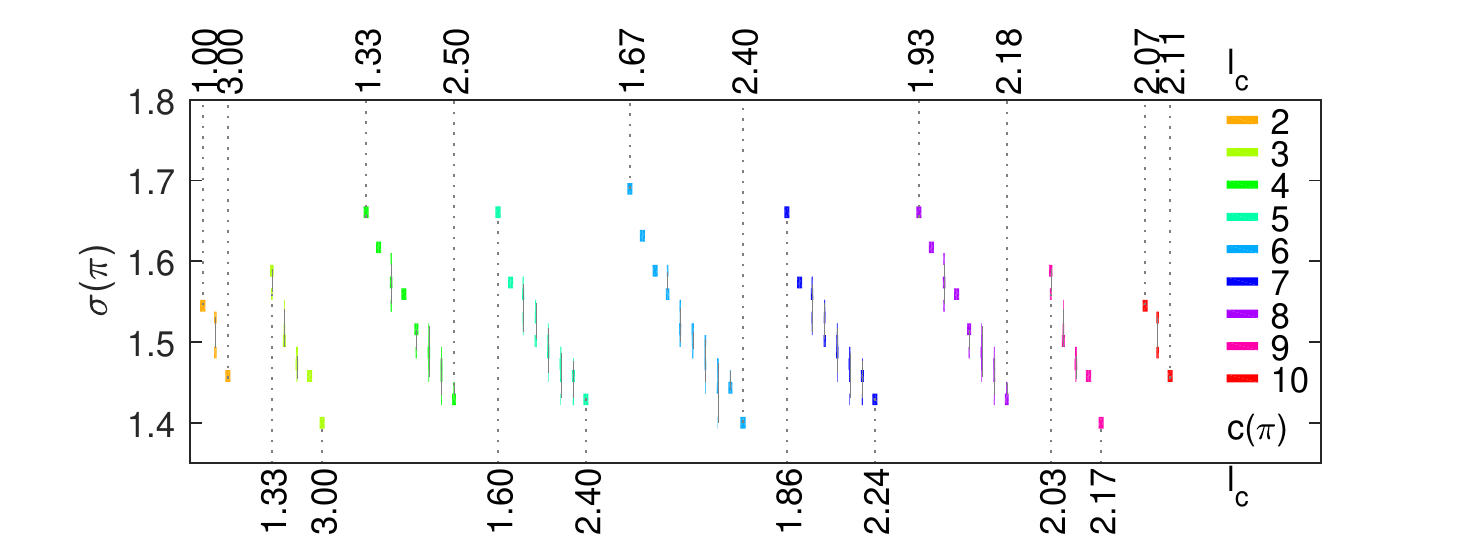}

\hspace{1cm} (f) M\"obius ladder graph, Fig. \ref{fig:loc_tietze_duer}(c)

\caption{Distribution of the structure coefficients $\sigma(\pi)$ as violin plots over unique values of the cooperator path length $l_c$ for each number of cooperators $2 \leq c(\pi) \leq 10$.}
\label{fig:viol_2}
\end{figure}

\clearpage

\begin{figure}[tb]

\includegraphics[trim = 0mm 0mm 0mm 0mm,clip,width=8.25cm, height=5.9cm]{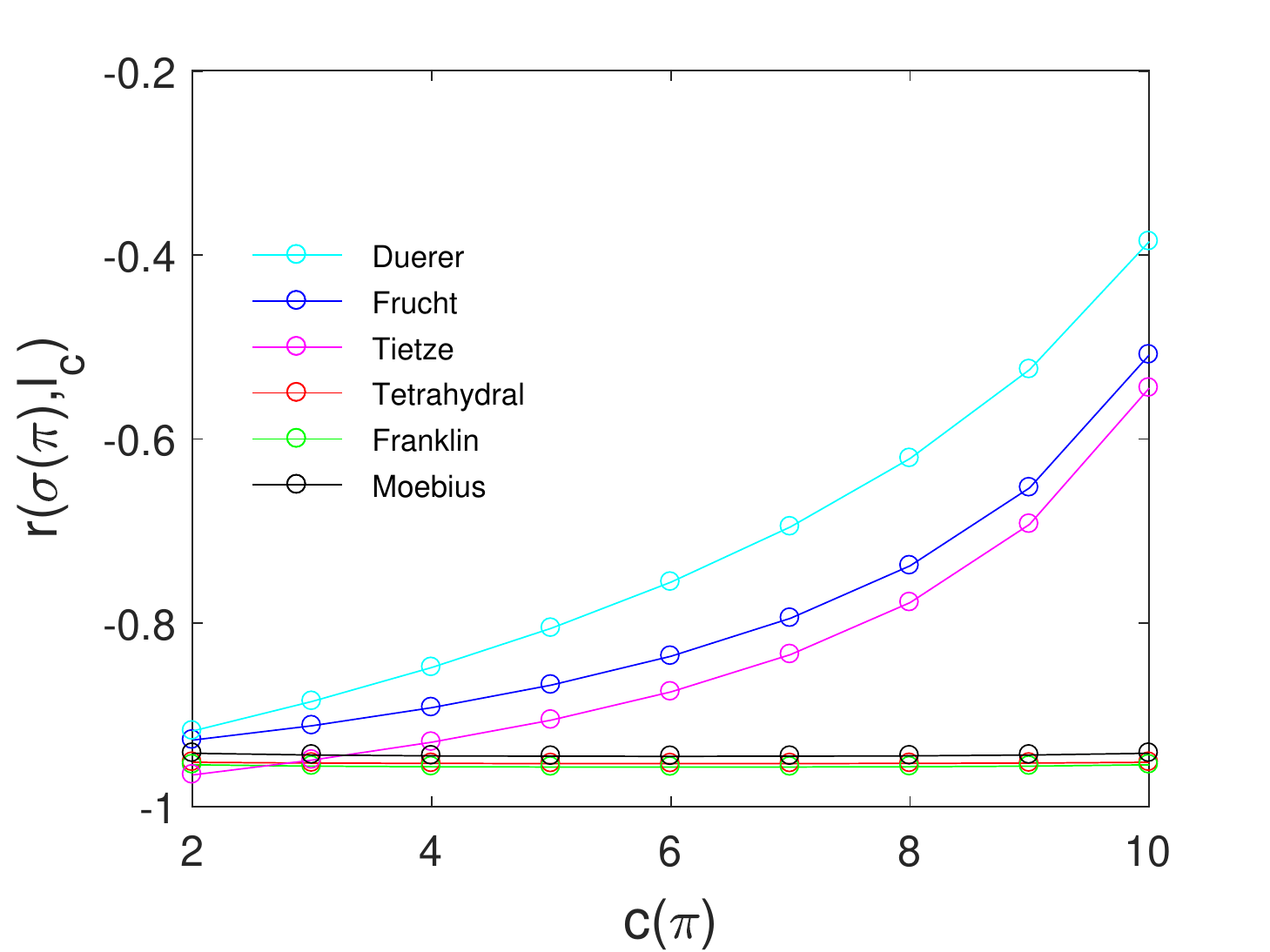}
\includegraphics[trim = 0mm 0mm 0mm 0mm,clip,width=8.25cm, height=5.9cm]{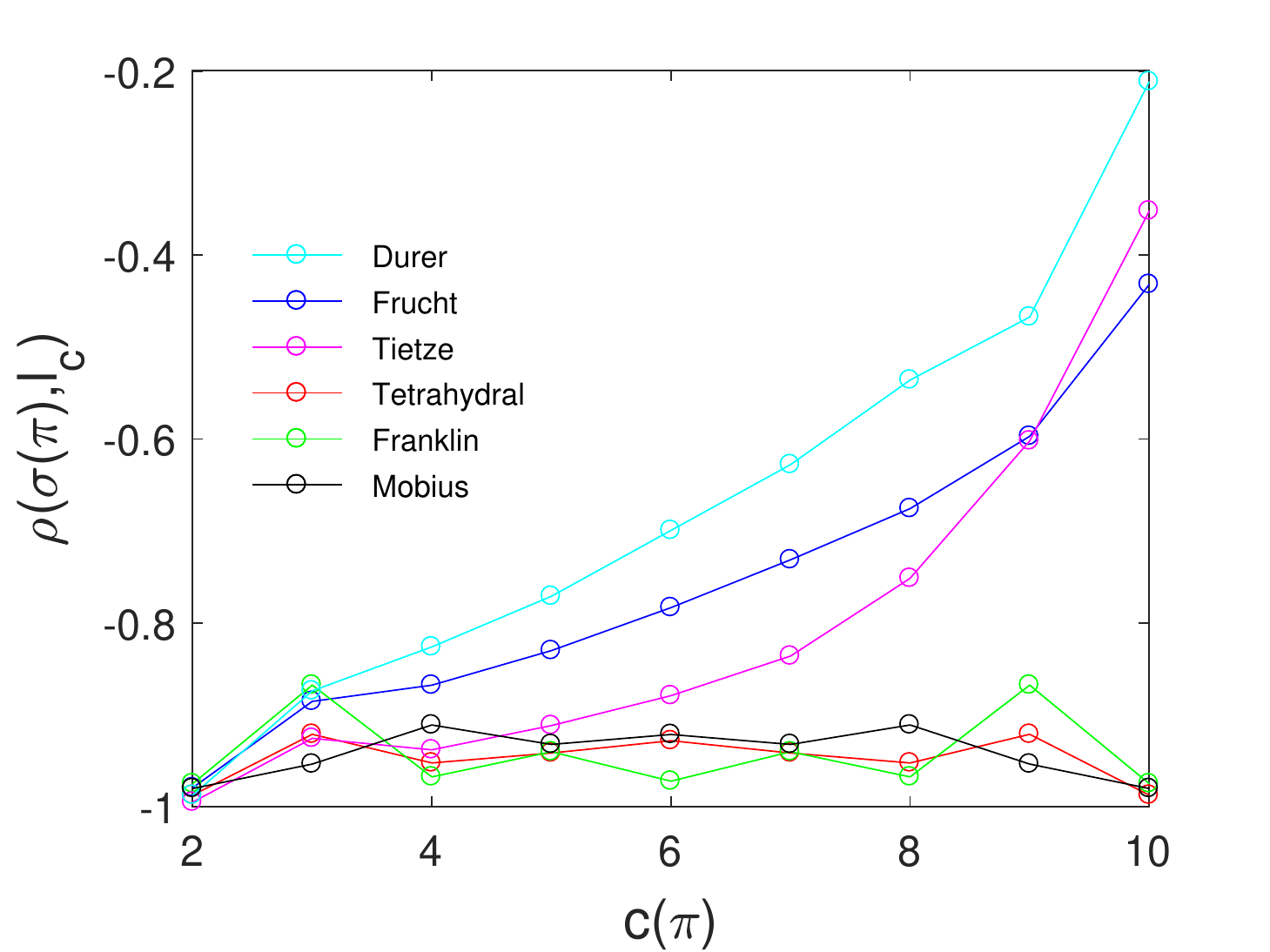}

\hspace{1cm} (a)  \hspace{7cm} (b) 

\vspace{2mm}

\caption{Correlation coefficients measuring how well the relationship between structure coefficients $\sigma(\pi)$ and cooperator path lengths $l_c$ can be described over the number of cooperators $c(\pi)$. (a) Pearson product--moment correlation coefficient $r$ giving a measure on a linear function. (b) Spearman rank correlation coefficient $\rho$ giving a measure on a monotonic function. }
\label{fig:corr_12_3_1}
\end{figure}

 As shown above for different evolutionary graphs, the values of $\sigma(\pi)$ (including the largest and smallest, but also intermediate values)  form a discrete distribution with $\#_\sigma$ different values. In other words, 
for every configuration $\pi$ we have a structure coefficient $\sigma(\pi)$ and a cooperator path length $l_c$. As the configurations can be classified according to the number of cooperators 
$c(\pi)$, we may study how the structure coefficients are distributed over $c(\pi)$ and also over the range of the cooperator path length $l_c$, see Figs.   \ref{fig:viol_frucht_frank}--\ref{fig:viol_2}. The distributions are depicted as histograms and shown as violin plots. The  number of bins of the histograms is calculated according to the Freedman--Diaconis rule~\cite{free81,scott92}, which  is designed to minimize the error made by the partition. The central axis of each violin plot gives the range between largest and smallest value of $\sigma(\pi)$. 

Fig. \ref{fig:viol_frucht_frank} shows some findings supporting the main results given with Fig. \ref{fig:viol_2} by considering the Frucht graph and the Franklin graph. 
In the upper panels, Fig. \ref{fig:viol_frucht_frank} (a),(b), the distributions of the structure coefficients $\sigma(\pi)$ are given for each number of cooperators, $1 \leq c(\pi) \leq 11$, while the lower panels, Fig. \ref{fig:viol_frucht_frank} (c),(d), take the example of 6 cooperators ($c(\pi)=6$) and depict the distributions for each unique value of   the  cooperator path length $l_c$. The results for the distribution of $\sigma(\pi)$ over the number of cooperators $c(\pi)$,  Fig. \ref{fig:viol_frucht_frank} (a),(b), show some characteristics that can similarly be found for other graphs as well.  
A first characteristics is that the distributions are symmetric with respect to the number of cooperators $c(\pi)$. A second is that for a single cooperator (or defector), $c(\pi)=1$ or $c(\pi)=N-1=11$, there is the same single value of the structure coefficient for all $N=12$ configurations,  see the blue line in Fig. \ref{fig:viol_frucht_frank}.  With $N=12$ and $k=3$ this value reproduces the well--known result $\sigma=\frac{(k+1)N-4k}{(k-1)N}=3/2$, compare with~\cite{lehmann07,nowak10,tarnita09,taylor07}. For $2 \leq c(\pi) \leq 10$ cooperators we obtain structure coefficients $\sigma(\pi)$ that are both larger and smaller than $\sigma=1.5$, depending on the arrangement of cooperators and defectors on the graph, see  Figs. \ref{fig:loc_frucht0} and \ref{fig:loc_tietze_duer} as well as the discussion in the Appendix 1. Put differently, for multiple cooperators the arrangement of cooperators and defectors on the evolutionary graph may substantially affect whether or not  cooperation is favored, even if the number of cooperators is the same.  Generally, the largest range between maximal and minimal value of $\sigma(\pi)$ is not reached for $c(\pi)=N/2$ but for $c(\pi)=N/2 \pm1$ or  $c(\pi)=N/2 \pm2$.

The distribution of $\sigma(\pi)$ over $c(\pi)$ for the Frucht graph, Fig. \ref{fig:viol_frucht_frank}(a), and the Franklin graph, Fig. \ref{fig:viol_frucht_frank}(b), are rather similar with the range between maximal and minimal value of $\sigma(\pi)$ slightly larger for the Frucht graph. However, comparing the results with respect to how the structure coefficients $\sigma(\pi)$ are distributed over unique values of the cooperator path length $l_c$ shows substantial differences between the graphs, see Fig. \ref{fig:viol_frucht_frank}(c),(d). Not only is the range of $l_c$ smaller for Franklin ($1.73 \leq l_c \leq 2.27$) than for Frucht ($1.60 \leq l_c \leq 2.40$), there are also fewer unique values ($\#_{l_c}=9$ vs. $\#_{l_c}=13$). Also, the range between $\sigma_{max}$ and $\sigma_{min}$ for each $l_c$ is smaller for Franklin than for Frucht.
For other numbers of cooperators, and the other interaction graphs considered, we find similar results. This applies in particular to the general result that large values of $\sigma(\pi)$ coincide very frequently with small values of $l_c$, and vice versa.  
To visualize how the histograms scale over the number of cooperators, Fig. \ref{fig:viol_2}
aggregates the violin plots over $c(\pi)$. Each violin is colored according to a color code for each $c(\pi)$, and on the axis below and above the graph the maximum and the minimum value of $l_c$ is given for each $c(\pi)$. 
By looking at this visual presentation of the relationships between structure coefficients $c(\pi)$ and cooperator path lengths $l_c$ several observations can be made. The first is that for small and medium numbers of cooperators $2 \leq c(\pi)\leq 6$ we see for all 6 graphs a clear reverse proportional relationship between $c(\pi)$ and $l_c$. For the vertex--transitive graphs this remains for $6 \leq c(\pi) \leq 10$. Moreover, for these graphs the distribution of $\sigma(\pi)$ for each value of $l_c$ shows a certain symmetry with respect to the number of cooperators. 
This is not true for the other 3 graphs. 
Given the fact that the whole distribution of $\sigma(\pi)$ over $c(\pi)$ is symmetric, see Fig. \ref{fig:viol_frucht_frank}(a),(b), this symmetry found for the 3 vertex--transitive graphs is caused by the cooperator path length $l_c$ showing certain symmetry properties.
 
 Compare the multiplicity $\#_{l_c}$ of $l_c$.
   For instance, for the Franklin  graphs we get the multiplicities of $l_c$ as $\#_{l_c}=\{3,3,9,5,9,5,9,3,3\}$ for the number of cooperators $c(\pi)=\{2,3,4,\ldots,9,10\}$. As for the number of cooperators $c(\pi)$ 
 and the number of defectors $d(\pi)$ there is $c(\pi)+d(\pi)=N$, the multiplicities  $\#_{l_c}$  for $c(\pi)$ matches that of $N-c(\pi)$. 
By contrast, for the Frucht graph we get $\#_{l_c}=\{4,7,9,12,13,14,13,11,8\}$, which does not show this kind of symmetry.    
  However, note that this symmetry does not mean that for each $c(\pi)$ and $N-c(\pi)$ we obtain the same values of the cooperator path length $l_c$. For instance, for the Franklin graph with $c(\pi)=2$ we have the values $l_c=\{1,2,3\}$, while for   $c(\pi)=10$ the values are   $l_c=\{1.98,2.00,2.02\}$. We have the same number of unique values but their range shrinks for $c(\pi)$ increasing. This appears plausible as for a small number of cooperators, for instance $c(\pi)=2$, the cooperator path length can be as high as the diameter of the graph, while for larger $c(\pi)$ the growing number of cooperators crowds the graph and makes average distances smaller.   This symmetry property of the distribution of $\#_{l_c}$ over $c(\pi)$, which can be found for the other 2 vertex transitives graphs as well, is related to the property that 
    vertex transitivity implies the graph to be strongly distance balanced~\cite{jere08,kutnar06,kutnar09}. 
    
We next analyze the relationships visualized in Fig. \ref{fig:viol_2} by considering correlation coefficients. Fig. \ref{fig:corr_12_3_1} shows the Pearson product--moment correlation coefficient $r$ and the Spearman rank correlation coefficient $\rho$, each between the variables $\sigma(\pi)$ and $l_c$. Pearson's $r$ gives a measure of how well the relationship between the variables can be described by a linear function, Fig. \ref{fig:corr_12_3_1}(a), while Spearman's $\rho$ is a measure of how well the relationship  matches a monotonic function, Fig. \ref{fig:corr_12_3_1}(b). We notice that for small $c(\pi)$ we obtain for both correlation coefficients and all 6 graphs values slightly bigger than $-1$, which indicates a strong inverse proportional relationship. For the 3 vertex--transitive graphs this strong correlation remains for all $c(\pi)$, while for the 3 graphs that are just regular, the correlation weakens while $c(\pi)$ increases. Comparing the two correlation coefficients shows that the relationships are better approximated by a linear than by a monotonic function.

\begin{figure}[tb]

\includegraphics[trim = 0mm 0mm 0mm 0mm,clip,width=8.25cm, height=5.9cm]{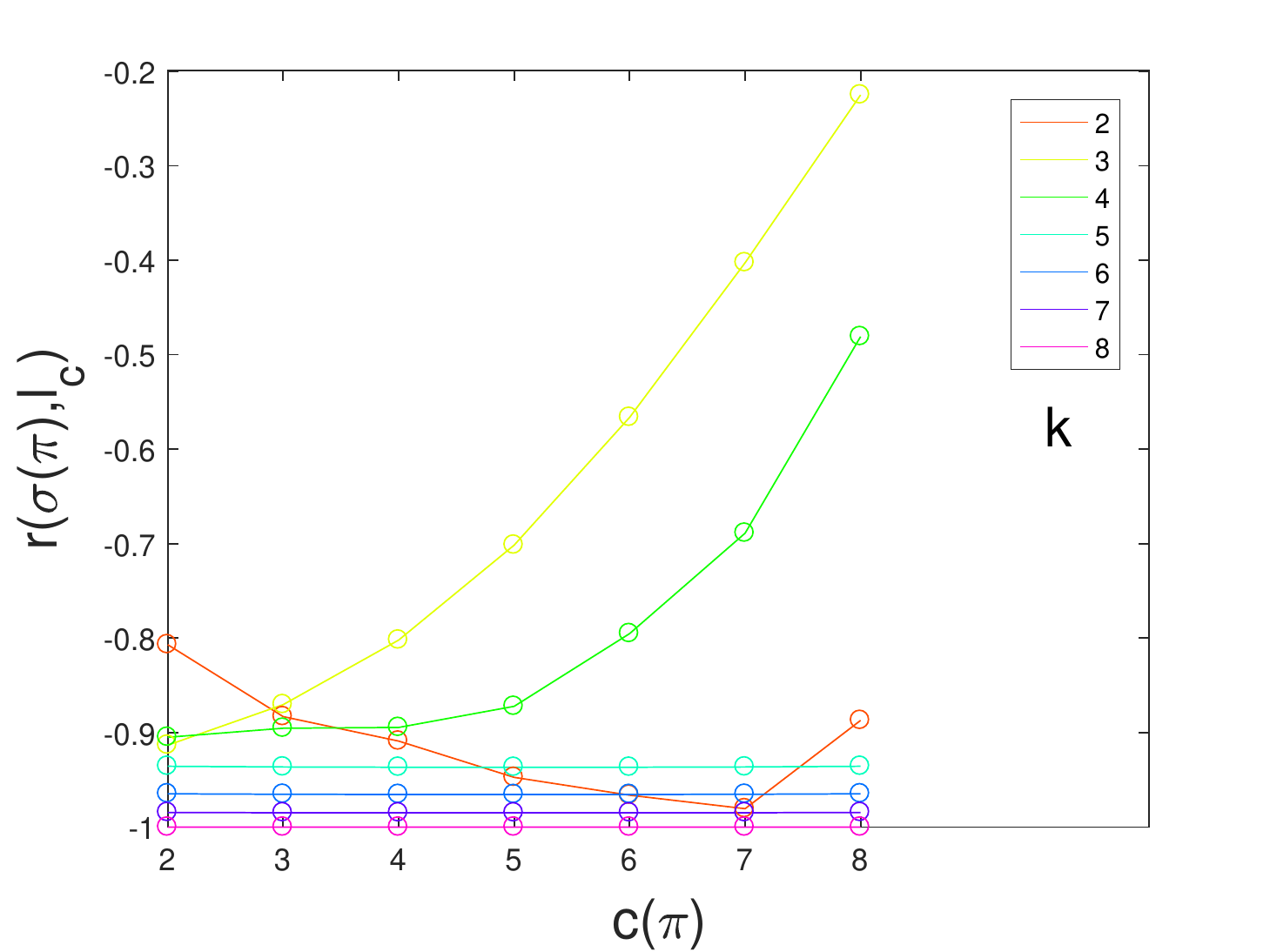}
\includegraphics[trim = 0mm 0mm 0mm 0mm,clip,width=8.25cm, height=5.9cm]{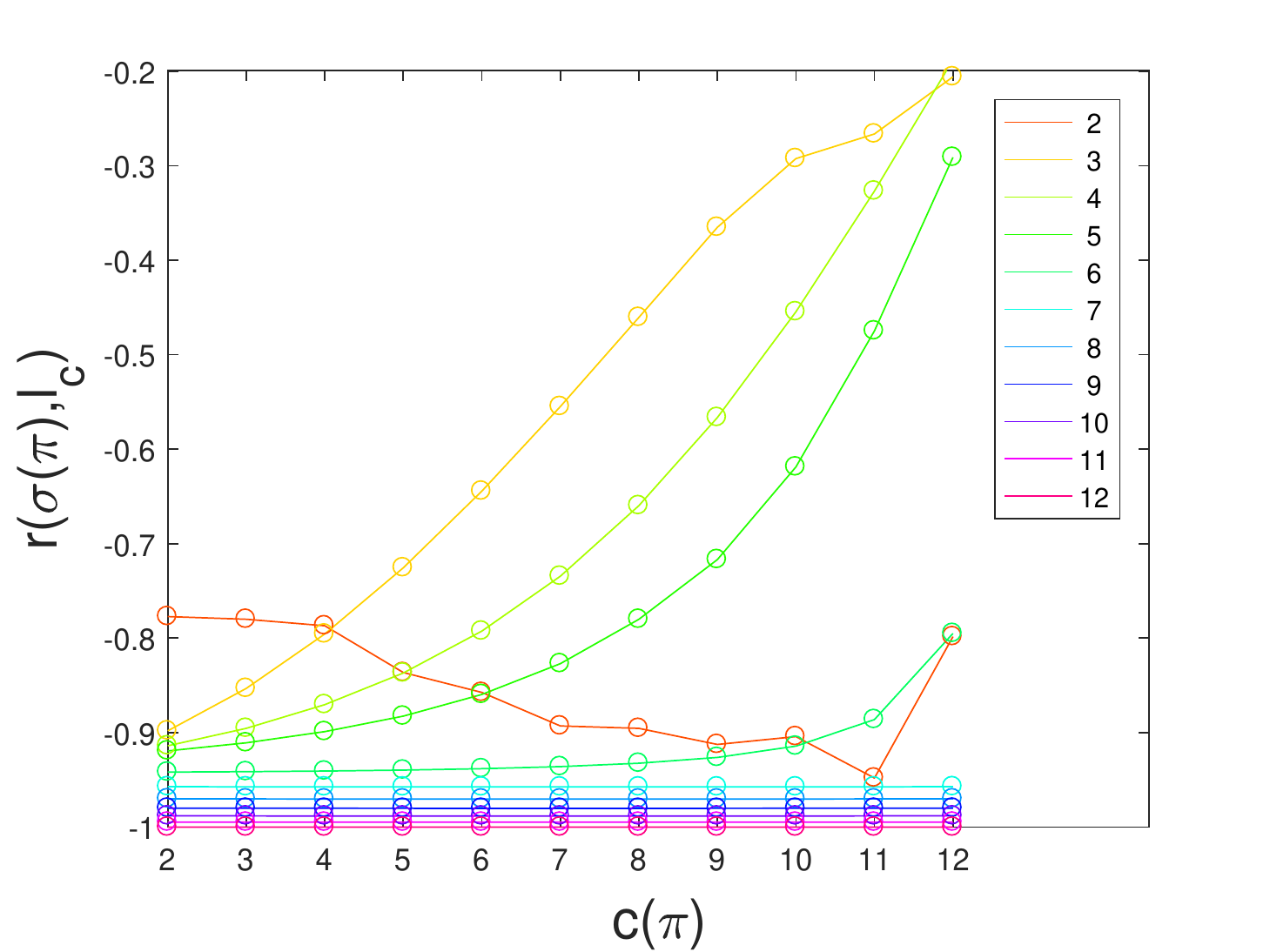}

\hspace{1cm} (a) $N=10$  \hspace{7cm} (b) $N=14$ 

\vspace{2mm}

\includegraphics[trim = 0mm 0mm 0mm 0mm,clip,width=8.25cm, height=5.9cm]{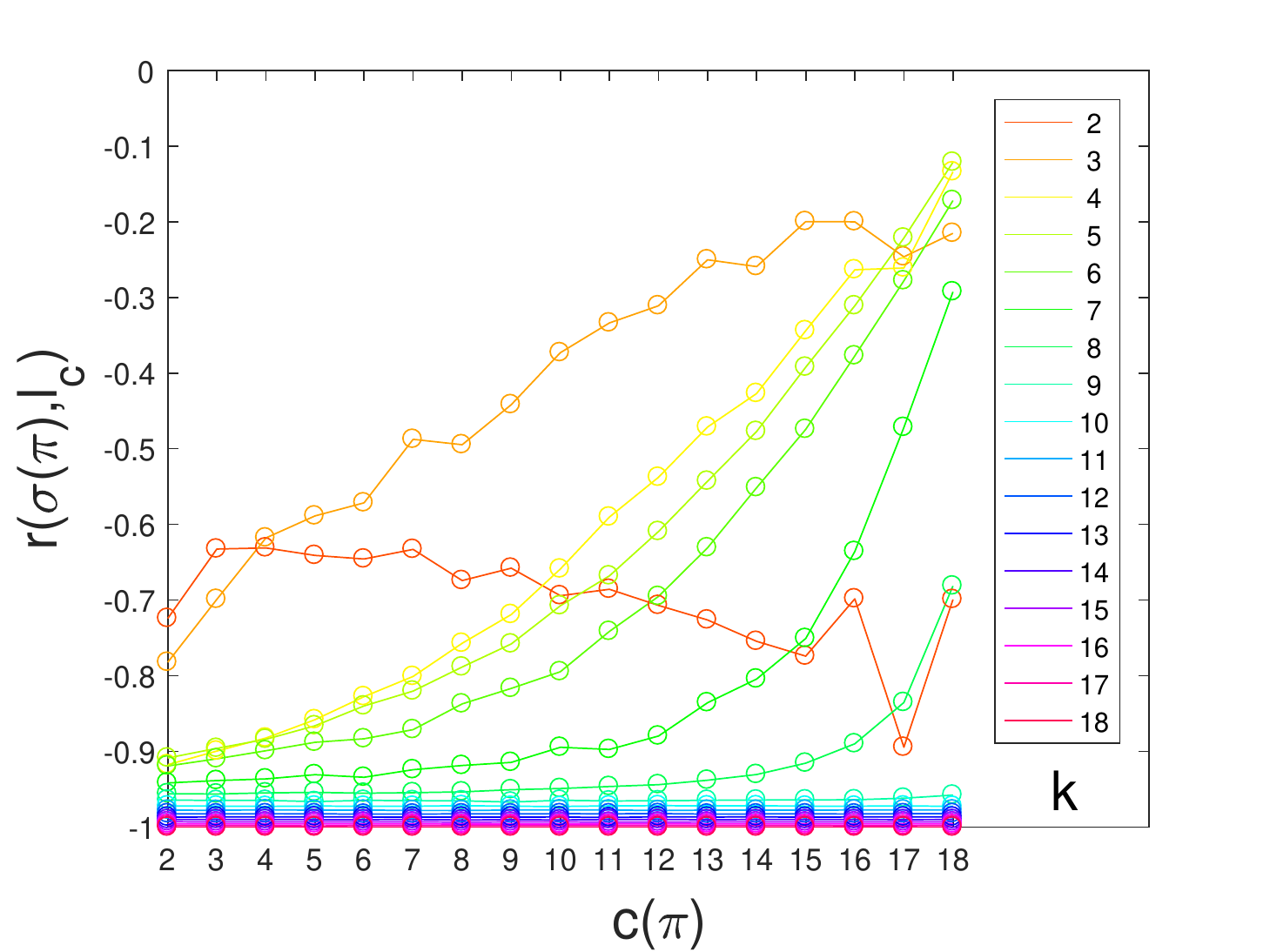}
\includegraphics[trim = 0mm 0mm 0mm 0mm,clip,width=8.25cm, height=5.9cm]{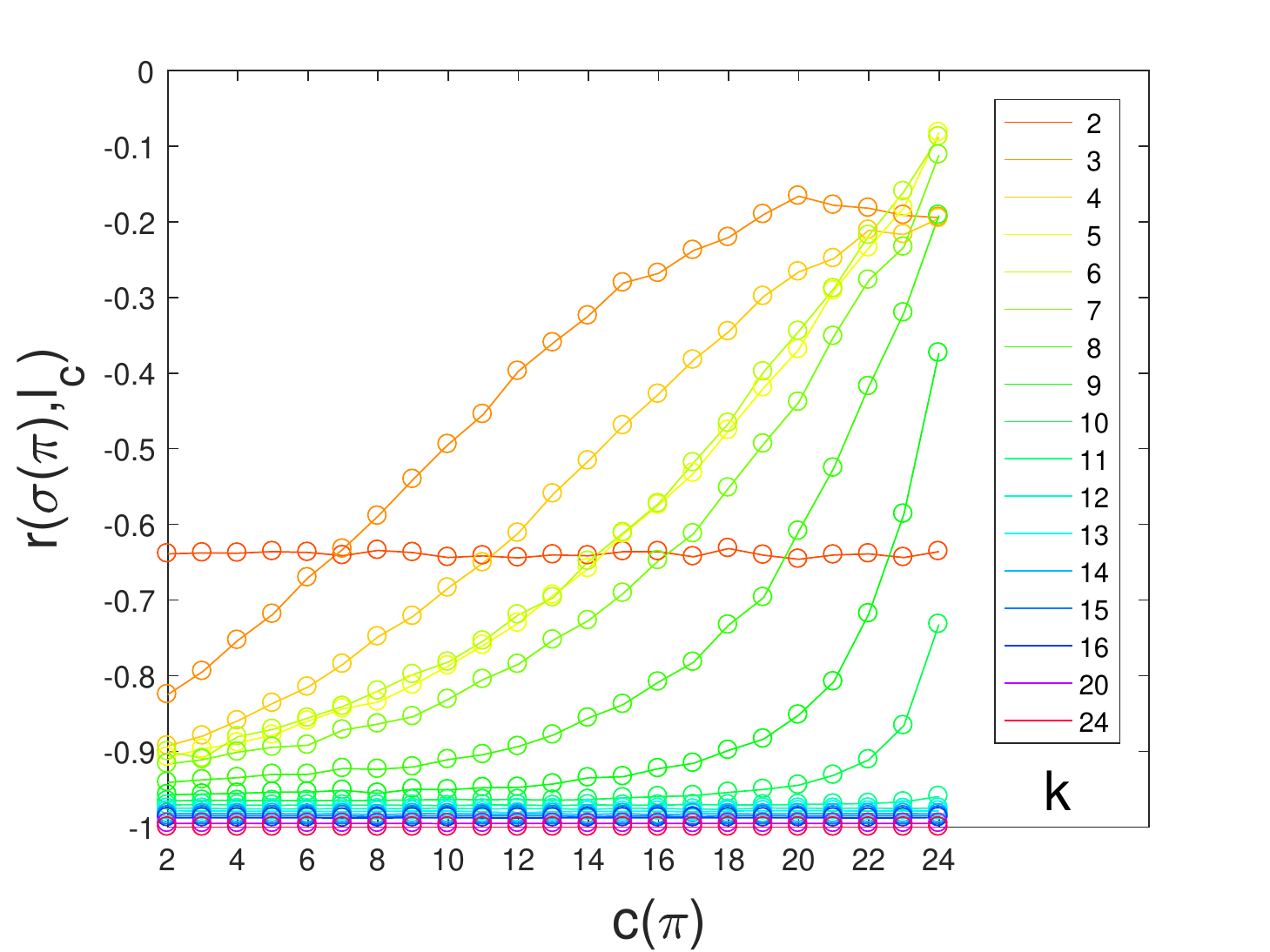}

\hspace{1cm} (c) $N=20$ \hspace{7cm} (d) $N=26$

\caption{ Pearson product--moment correlation coefficient $r$ measuring how well the relationships between structure coefficients $\sigma(\pi)$ and cooperator path lengths $l_c$ can be described by  linear functions over the number of cooperators $c(\pi)$ for $N=\{10,14,20,26 \}$ players and all $2 \leq k \leq N-1$ coplayers.  }
\label{fig:corr_samp_compl}
\end{figure} 

\clearpage

We close with a
 brief study of how the correlation between $\sigma(\pi)$ and $l_c$ behaves for other $N$ and $k$. Therefore, we analyze coevolutionary games with $N=\{10,14,20,26 \}$ players, each for all $2 \leq k \leq N-1$ coplayers. Interaction networks are modeled as regular graphs, whose adjacency matrices $A_I$ are generated algorithmically~\cite{bay10,blitz11}. We discard graphs that are not connected and vertex--transitive, which is based on analyzing the Fiedler eigenvalue (the second--smallest eigenvalue of the Laplacian matrix $kI-A_I$)  for connectivity~\cite{brou12}, and  symmetry properties of  transition matrices describing  random walks on $A_I$ with Markov chain characteristics  for vertex transitivity~\cite{sli86}. For each $N$ and $k$, we use a set of $1000$ different graphs. For $N=\{10,14\}$,  all $2^N$ configurations are analyzed. As analyzing all configurations   could not be realized with the computational resources available in this study for  $N=\{20,26\}$, we sampled $25.000$ configurations uniformly at random out of the whole set. Auxiliary results with different samples  have shown that the error made in calculating the correlations by such a sampling is very small, which is plausible as there is an increasing number of  configurations with the same $\sigma(\pi)$ and $l_c$.
Fig. \ref{fig:corr_samp_compl} shows the correlations expressed as  Pearson's $r$, which are averages over all 1000 graphs. We see that the results generally support the claims made for the 6 graphs with $N=12$ and $k=3$. Particularity for $3 \leq k \leq N/2$ coplayers the curves resemble those for the Frucht, the D\"urer and the Tietze graph, see Fig. \ref{fig:corr_12_3_1}(a). For $k=2$  (cycle graphs), the strength of correlation increases with increasing $c(\pi)$ (except for $N=26$), while for $k \geq N/2$ the correlations remain strong for an increasing number of cooperators, similar to the results for the vertex--transitive graphs. The results in Fig. \ref{fig:corr_samp_compl} suggest that the correlations reported are not specific for the 6 graphs listed in Tab. \ref{tab:graph}, but apply more generally. Further works could study if this is also the case for $N>26$ players.

\section{Discussion}
Whether or not cooperation is favored over defection in an evolutionary game on $k$--regular graphs with $N$ vertices ($=$ players) can be assigned in the case of weak selection by the structure coefficient $\sigma(\pi)$ for any arrangement of cooperators and defectors described by a configuration $\pi$~\cite{chen16,rich18}. As the emergence of cooperation is always opposed for BD updating, the focus here was on the apparently more interesting case of DB updating. For DB updating,  $\sigma(\pi)$ can be calculated with polynomial time complexity by Eq.~(\ref{eq:sigma}). It was illustrated using the example of 6 different interaction networks modeled as regular graphs that for 2 cooperators (and up to $N-2$ cooperators), the structure coefficients may vary over configurations, even if the number of cooperators is the same. Thus, if we interpret a single cooperator among defectors as a lone mutant, the results reported apply to configurations of multiple mutants as well. Furthermore, numerical results were given suggesting that the structure coefficients $\sigma(\pi)$ can be linked to the cooperator path length $l_c$ by an inverse linear relationship. Additional results with a multitude of interaction graphs algorithmically generated with  $N=\{10,14,20,26 \}$ players and  $2 \leq k \leq N-1$ coplayers confirm these findings. As this implies large values of $\sigma(\pi)$ for  low cooperator path lengths $l_c$, and vice versa, the result provides another rationale for  clusters of cooperators giving different fixation properties than cooperators that are sparsely distributed on the evolutionary graph. For evolutionary games on lattice grids  clusters of cooperators have better chances to survive~\cite{hau01,hau04,lang08,page00}. Lattice grids can be understood as $4$--regular graphs for Von Neumann neighborhoods and as $8$--regular graphs for Moore neighborhoods. Clusters of cooperators have smaller cooperator path lengths than cooperators that are isolated. Thus, the results for lattice grids are consistent with (and a special case of) the findings for $k$--regular graphs given here.   
 
The approach presented in this paper gives a methodology for analyzing how properties of interaction networks relate to arrangements of more than one cooperator and subsequently to fixation properties of the evolutionary game specified by the graph. It may add    a new perspective at finding best conditions to favor cooperation. Within the framework and interpretation of evolutionary graph theory given in Sec.~\ref{sec:methods}, this means looking for the best arrangement of a given number of cooperators on a given interaction graph, but possibly also for the best interaction graph with a given number and arrangement of cooperators, or the best combination of number, arrangement and graph.   With the payoff matrix (\ref{eq:payoff}) and the structure coefficient $\sigma(\pi)$ cooperation is favored if \begin{equation} \sigma(\pi)>\frac{c-b}{a-d}. \end{equation}
Note that this condition indicates that the fixation probability of cooperation is higher than  the fixation probability of defection, but entails no statement about the exact values of the fixation probabilities themselves. As known for the Frucht and for the Tietze graph, for a single cooperator the structure coefficient does not imply the exact value of the fixation probability of cooperation. For a single cooperator we have a single value of the structure coefficient but different fixation probabilities which may vary over initial configurations~\cite{mcavoy15}.

 To refocus on whether or not cooperation is favored, consider, for example, the parametrization of the payoff matrix (\ref{eq:payoff})  used in the seminal work of Axelrod~\cite{axel80}, with later modifications by Nowak \& Sigmund~\cite{nowak93}, for the Prisoner's dilemma (PD) and the snowdrift (SD) game: $ \begin{pmatrix} a &  b \\ c & d \end{pmatrix}= \begin{pmatrix} 4 &  0 \\ 5 & 1 \end{pmatrix}$ for PD and $ \begin{pmatrix} 3 &  1 \\ 5 & 0 \end{pmatrix}$ for SD. We obtain that cooperation is favored over defection if $\sigma(\pi)>5/3$ for the PD and  if $\sigma(\pi)>4/3$ for the SD game. In view of the findings given in Sec.~\ref{sec:results}, this means that 2 cooperators  attain fixation on none of the 6 graphs considered for the PD game, but on all 6 graphs for the SD game. For $c(\pi)=2$, all $\sigma_{max}<5/3$, and all $\sigma_{min}>4/3$. For a single cooperator the same applied as $\sigma=3/2$. Consequently, for the SD game fixation of cooperation occurs for more than 2 cooperators as well. For the PD game and $c(\pi) \geq 3$  cooperators, the situation is different. As $\sigma_{min}<5/3$ for all 6 graphs and all $c(\pi)$  for the interaction networks considered, see Tab. \ref{tab:graph}, cooperation is never unconditionally favored, that is for all configurations and independent of how many cooperators $c(\pi) \geq 3$ there are. However, except for the Franklin graph there are configurations that favor cooperation with $\sigma(\pi)>5/3$,  see Fig. \ref{fig:viol_frucht_frank}. For instance, for the Frucht graph and $c(\pi)=3$ cooperators, the configuration  shown in Fig. \ref{fig:loc_frucht0}(a)  has $\sigma(\pi)=1.6897>5/3$. The same applies to the other 2 configurations occupying the other 2 triangles of the Frucht graph. Similarly, for the other graphs in Figs. \ref{fig:loc_frucht0} and \ref{fig:loc_tietze_duer} (see also the Appendix 1)  as well as certain numbers of cooperators $c(\pi)$, there are some  configurations $\pi$ favoring cooperation. Comparing the graphs  suggests again a difference between the vertex--transitive graphs and those that are just regular with the 3 vertex--transitive graphs considered more rarely favoring cooperation with the parameters of the payoff matrix given.  Whether this result is generally true for vertex--transitive graphs or 
an anomaly due to special properties of the truncated tetrahedral, the Franklin and the M\"obius ladder graph should be addressed by future work. Thus, if we were  just to consider the 6 interaction networks listed in Tab. \ref{tab:graph}, the Frucht graph with $c(\pi)=5$ cooperators and configuration $\pi=(1000\:0000\:1111)$ would be most favorable for cooperation as this combination of network and configuration yields  the largest value $\sigma_{max}=\sigma(1000\:0000\:1111)=1.7568$. Due to the symmetry of $\sigma(\pi)$ with respect to $c(\pi)$ another solution is the configuration $\pi=(0111\:1111\:0000)$ with $c(\pi)=7$ and the same $\sigma_{max}=\sigma(0111\:1111\:0000)=1.7568$. In other words, if we view the parameter $c=5$ of the payoff matrix of the PD game as the temptation to defect, then these 2 configurations  would  favor cooperation on the Frucht graph even if the temptation gets as high as $c\leq 3 \sigma_{max}=5.2704>5$. Moreover,  for the PD game with $a=4$ the reward to cooperate and $d=1$ the punishment to defect, cooperation is Pareto--efficient as $a-d=3>0$. Thus, the 2 configurations on the Frucht graph would favor cooperation even if  the value of the difference between reward and punishment is as low as $a-d \geq 5/\sigma_{max}=2.8461<3$. In  a similar manner the effect of configuration and network on relations between the parameters of the payoff matrix can be studied, for instance on risk--dominance $(a+b)/(c+d)$.

Another interesting aspect and pointer to future work comes
from analyzing properties of the configurations with largest $\sigma(\pi)$. We notice that $\pi=(1000\:0000\:1111)$ indicates that the 5 cooperators are situated on the overlap of the square and a triangle of the Frucht graph, while the 7 cooperators of    $\pi=(0111\:1111\:0000)$ are on the overlap of 2 triangles and the hexagon (cycle of length 6).  These results relate to findings showing for 3 and 4 cooperators that the largest structure coefficients $\sigma(\pi)$ are frequently found for the cooperators located on triangles or squares of the evolutionary graph, see for instance Fig. \ref{fig:loc_frucht0}.  Naturally, if the cooperators are located on these cycles, such overlaps of cycles with different lengths coincide with low values of the  cooperator  path length. 
These results suggest the hypothesis that particularly large values of $\sigma(\pi)$ can be obtained if the interaction graph possesses a certain abundance of cycles of length up to the number of cooperators $c(\pi)$. Large values of $\sigma(\pi)$ appear if the cooperators are located on these cycles, particularly if the cycles overlap, that is  cycles of different lengths share vertices.
Further work is needed for clarification.

This kind of discussion immediately leads to a related question, which is to design optimal conditions of cooperation. Freezing the values of the payoff matrix (\ref{eq:payoff}), the design space encompasses the initial configurations as well as the interaction networks. 
For games with 2 strategies and $N$ players the number of configurations is $2^N$ and just considering evolutionary graphs that are regular their number grows for $k=o(\sqrt{N})$ with  $\mathcal{O}(N^N)$~\cite{rich17,worm99}. Thus, while for a small number of players, such as $N=12$ as in the examples discussed, and  pre--selected interaction networks as in Tab. \ref{tab:graph}, a calculation of all $\sigma(\pi)$ can be done,  solving the design problem in general becomes numerically expensive for intermediate $N$ and infeasible for $N$ getting large.  For designing configurations, there are, however, some further aspects.  For $N \rightarrow \infty$, the structure coefficients uniformly tend to $\sigma(\pi)\rightarrow \sigma =(k+1)/(k-1)$~\cite{chen16,ohts06}. 
Put differently, whether cooperation is favored ceases to depend on which configuration the game has initially and thus solving the design problem ceases to be particularity meaningful  for the number of players $N$ getting large.
Another aspect is the scaling of the number of configurations $\#_{c(\pi)}$ over the number of cooperators $c(\pi)$, which is governed by Eq.  (\ref{eq:config_number}). Accordingly, $\#_{c(\pi)}$ grows polynomially with $c(\pi)$ for all $c(\pi) \neq N/2$, for instance by   $\#_{2}=(N-1)N/2$ for $c(\pi)=2$ or by $\#_{3}=(N-2)(N-1)N/6$ for $c(\pi)=3$. Only for $c(\pi)=N/2$ and $N \rightarrow \infty$, $\#_{N/2}$ grows exponentially. Thus, from a numerical point of view designing an optimal configuration for a given number of cooperators should remain feasible, except for $c(\pi)=N/2$, at least for $N$ not as large as the structure coefficients becoming uniform with $\sigma(\pi) \approx \sigma \approx (k+1)/(k-1)$.
 The results also have the potential to inform finding interaction graphs that are better suited than a random selection to either promote or suppress cooperation. For instance, the Franklin graph completely suppresses  cooperation in the PD game discussed above, while the Frucht graph  promotes cooperation much more. The analysis suggests that graphs with overlapping cycles of length up to the number of cooperators produce more favorable  structure coefficients than graphs with few or separate cycles. 
Thus, we may potentially prescribe interaction networks with these  properties and thus reduce the size of the design space.

The results given are for evolutionary game dynamics with weak selection and interaction networks modeled by  regular graphs. The paper closes with brief comments on going beyond
regular graphs and weak selection.
 Recently, an approach to assign fixation properties using coalescence times of random walks has been proposed that is suitable for a single cooperator and any graph structure~\cite{allen17}. If this technique can be extended to configurations with more than one cooperator, the approach given here would become applicable for graphs that are not regular. 
It is known that the validity of extrapolating results from weak to  intermediate and strong selection is not always possible and depends highly on game characteristics, population size and spatial heterogeneity of the network~\cite{fu09,mull14,wu13}, but comparison between fixation probabilities has been shown to be fairly robust for a varying intensity of selection and a single cooperator~\cite{fu09}. It remains to be seen if this is also valid for any arrangement of cooperators and defectors.

\section*{Appendix 1}
A more detailed description of the relationships between  configurations $\pi$, structure coefficients $\sigma(\pi)$ and cooperator path lengths $l_c$ is presented for selected graphs.  

\begin{figure}
\includegraphics[trim = 10mm 80mm 0mm 80mm,clip,width=17.25cm, height=11.25cm]{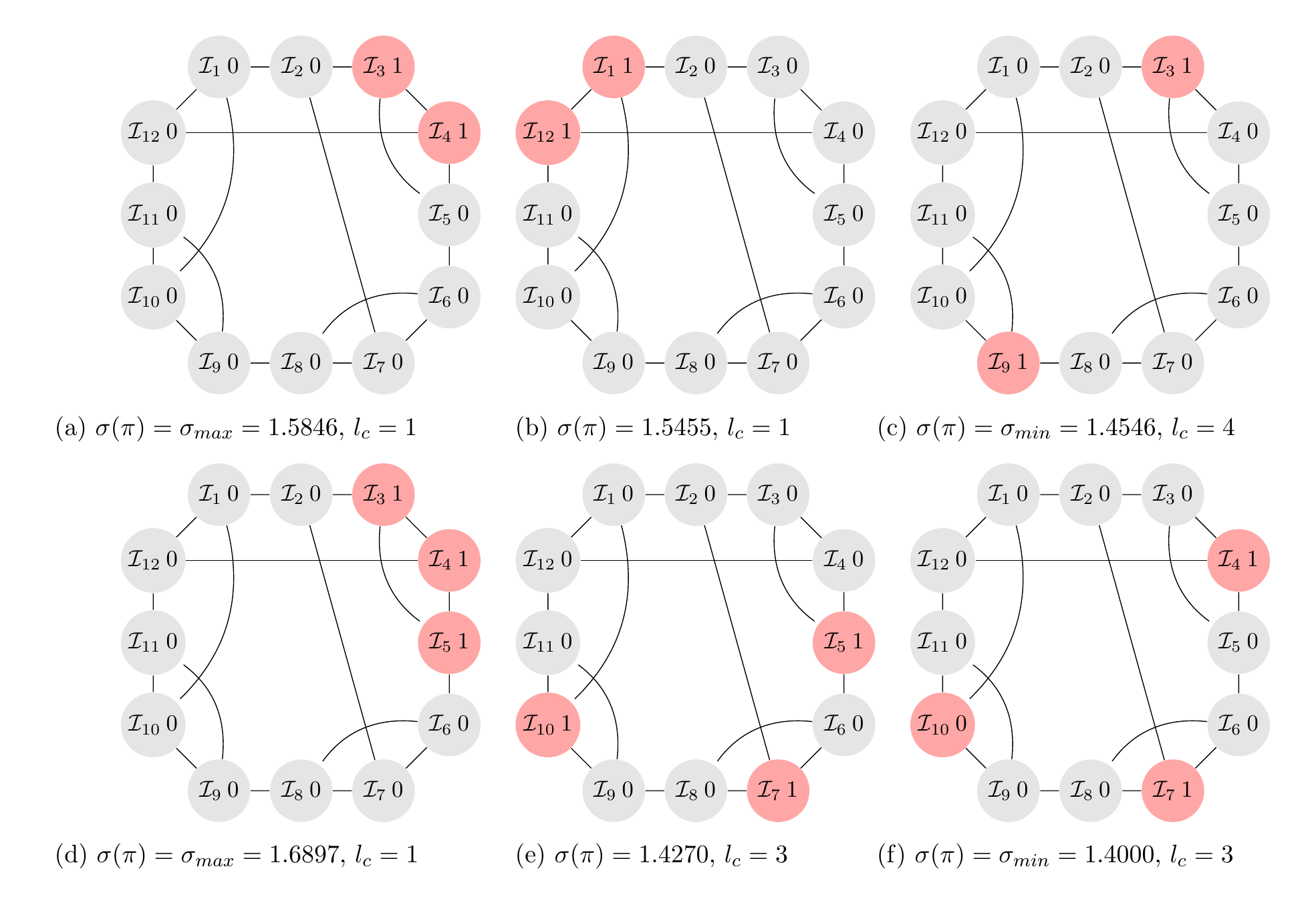}

\caption{\small{Configurations $\pi$, structure coefficients $\sigma(\pi)$ and cooperator path lengths $l_c$ for the Frucht graph~\cite{chen16,fru49,mcavoy15}, which has no non--trivial symmetry. (a)--(c):  Configurations with 2 cooperators ($c(\pi)=2$). 
(d)--(f): Configurations with 3 cooperators ($c(\pi)=3$).
}}
\label{fig:loc_fre}
\end{figure}

 1. The Frucht graph~\cite{chen16,fru49,mcavoy15}, which has no non--trivial symmetry. Fig. \ref{fig:loc_fre} (a)--(c):  Configurations with 2 cooperators ($c(\pi)=2$. 
There are  $\#_2=66$ configurations with  2 cooperators according to Eq. (\ref{eq:config_number}). Calculating the structure coefficients $\sigma(\pi)$ by Eq. (\ref{eq:sigma}) yields  $\#_{\sigma_{max}}=9$ configurations out of these 66 that have the maximal value of $\sigma_{max}=1.5846$. All these configurations have the 2 cooperators distanced by the minimal cooperator path length $l_c=1$ \emph{and} the 2 cooperators belonging to one of the 3 triangles of the Frucht graph $(\mathcal{I}_3,\mathcal{I}_4,\mathcal{I}_5)$, $(\mathcal{I}_6,\mathcal{I}_7,\mathcal{I}_8)$ and $(\mathcal{I}_9,\mathcal{I}_{10},\mathcal{I}_{11})$, see Fig. \ref{fig:loc_fre}(a) showing the example of configuration $\pi=(0011\:0000\:0000)$. The minimal cooperator path length  $l_c=1$ alone also yields the second largest $\sigma(\pi)=1.5455$, but $l_c=1$ is not sufficient for the maximal value, see Fig. \ref{fig:loc_fre}(b) showing the configuration $\pi=(1000\:0000\:0001)$. The smallest value of $\sigma_{min}=1.4546$ corresponds with the $\#_{\sigma_{min}}=23$ configurations with largest values of  $l_c$, (21 configurations with $l_c=3$ and 2 configurations with $l_c=4$), see  the example of $\pi=(0010\:0000\:1000)$ with $l_c=4$ in Fig. \ref{fig:loc_fre}(c). Over all configurations with $c(\pi)=2$, there are $\#_\sigma=5$ different values of $\sigma(\pi)$. Fig. \ref{fig:loc_fre} (d)--(f): Configurations with 3 cooperators ($c(\pi)=3$). The largest value $\sigma_{max}=1.6897$ is obtained for $\#_{\sigma_{max}}=3$ configurations out of the $\#_3=220$. Configurations maximizing $\sigma(\pi)$ are characterized by the minimal $l_c$ such that the cooperators are occupying the 3 triangle of Frucht graph, see  Fig. \ref{fig:loc_fre}(d) showing the example $\pi=(0011\:1000\:0000)$ which has $l_c=1$ according to Eq. (\ref{eq:av_path}) with individual path lengths $d=(d_{34},d_{35},d_{45})=(1,1,1)$. Small values of $\sigma$ are obtained for large values of $l_c$. There are 2 configurations with the largest value $l_c=3$ for which $\sigma(\pi)=1.4270$ (the second smallest value) is obtained for individual path lengths between cooperators $d=(d_{57},d_{5 \:10},d_{7 \: 10})=(2,4,3)$, Fig.\ref{fig:loc_fre}(e), while the smallest value of $\sigma_{min}=1.4000$  belongs to the configuration $\pi=(0001\:0010\:0100)$ with  $d=(d_{47},d_{4 \:10},d_{7 \: 10})=(3,3,3)$, Fig. \ref{fig:loc_fre}(f).   Over all configurations with $c(\pi)=3$, there are $\#_\sigma=14$ different values of $\sigma(\pi)$.

\begin{figure}

\includegraphics[trim = 10mm 80mm 0mm 80mm,clip,width=17.25cm, height=11.25cm]{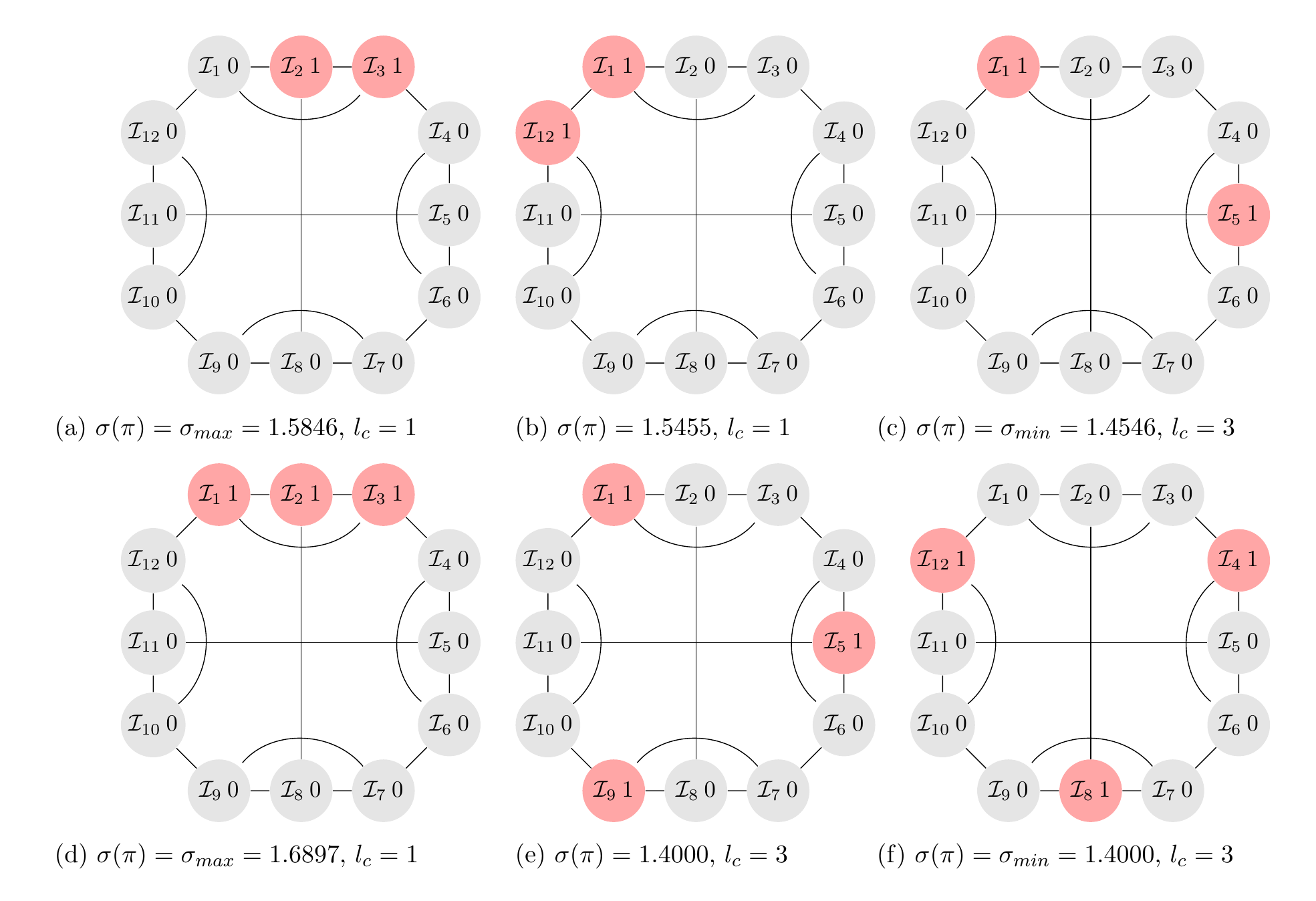}
\caption{\small{Configurations $\pi$, structure coefficients $\sigma(\pi)$ and cooperator path lengths $l_c$ for the truncated tetrahedral graph~\cite{read98}, which is vertex--transitive and square free. (a)--(c):  Configurations with 2 cooperators ($c(\pi)=2$).   (d)--(f):  Configurations with 3 cooperators ($c(\pi)=3$). 
}}
\label{fig:loc_tetra}
\end{figure}

2. The truncated tetrahedral graph~\cite{read98}, which is vertex--transitive and square free. Fig. \ref{fig:loc_tetra}(a)--(c):  Configurations with 2 cooperators ($c(\pi)=2$).   There are $\#_{\sigma_{max}}=12$ out of 66 configurations that have the maximal $\sigma_{max}=1.5846$, all of which have the minimal $\l_c=1$ \emph{and} belong to one of the 4 triangles of the  truncated tetrahedral graph $(\mathcal{I}_1,\mathcal{I}_2,\mathcal{I}_3)$, $(\mathcal{I}_4,\mathcal{I}_5,\mathcal{I}_6)$,  $(\mathcal{I}_7,\mathcal{I}_8,\mathcal{I}_9)$ and $(\mathcal{I}_{10},\mathcal{I}_{11},\mathcal{I}_{12})$, see Fig.~\ref{fig:loc_tetra}(a) showing the example of configuration $\pi=(0110\:0000\:0000)$. As for the Frucht graph (Fig. \ref{fig:loc_fre}), the minimal  $l_c=1$ alone also yields the second highest value $\sigma(\pi)=1.5455$, see Fig. \ref{fig:loc_tetra}(b). There are 24 configurations  with the maximal $l_c=3$  which all give the smallest value of $\sigma_{min}=1.4546$, see Fig. \ref{fig:loc_tetra}(c). Over all configurations with $c(\pi)=2$, there are $\#_\sigma= 4$ different values of $\sigma(\pi)$.  Fig. \ref{fig:loc_tetra}(d)--(f):  Configurations with 3 cooperators ($c(\pi)=3$). The largest value $\sigma_{max}=1.6897$ is obtained for 4  out of 220 configurations, each configuration representing one of the 4 triangles of the  graph, see the example  $\pi=(1110\:0000\:0000)$ in Fig. \ref{fig:loc_tetra}(d). The smallest value $\sigma_{min}=1.4000$ is obtained for the 4 configurations with $l_c=3$, see Figs.\ref{fig:loc_tetra}(e) and (f) with the examples $\pi=(1000\:1000\:1000)$ and $\pi=(0001\:0001\:0001)$. There are  $\#_\sigma=10$ different values of  $\sigma(\pi)$ with $c(\pi)=3$.

\begin{figure}
\includegraphics[trim = 10mm 80mm 0mm 80mm,clip,width=17.25cm, height=11.25cm]{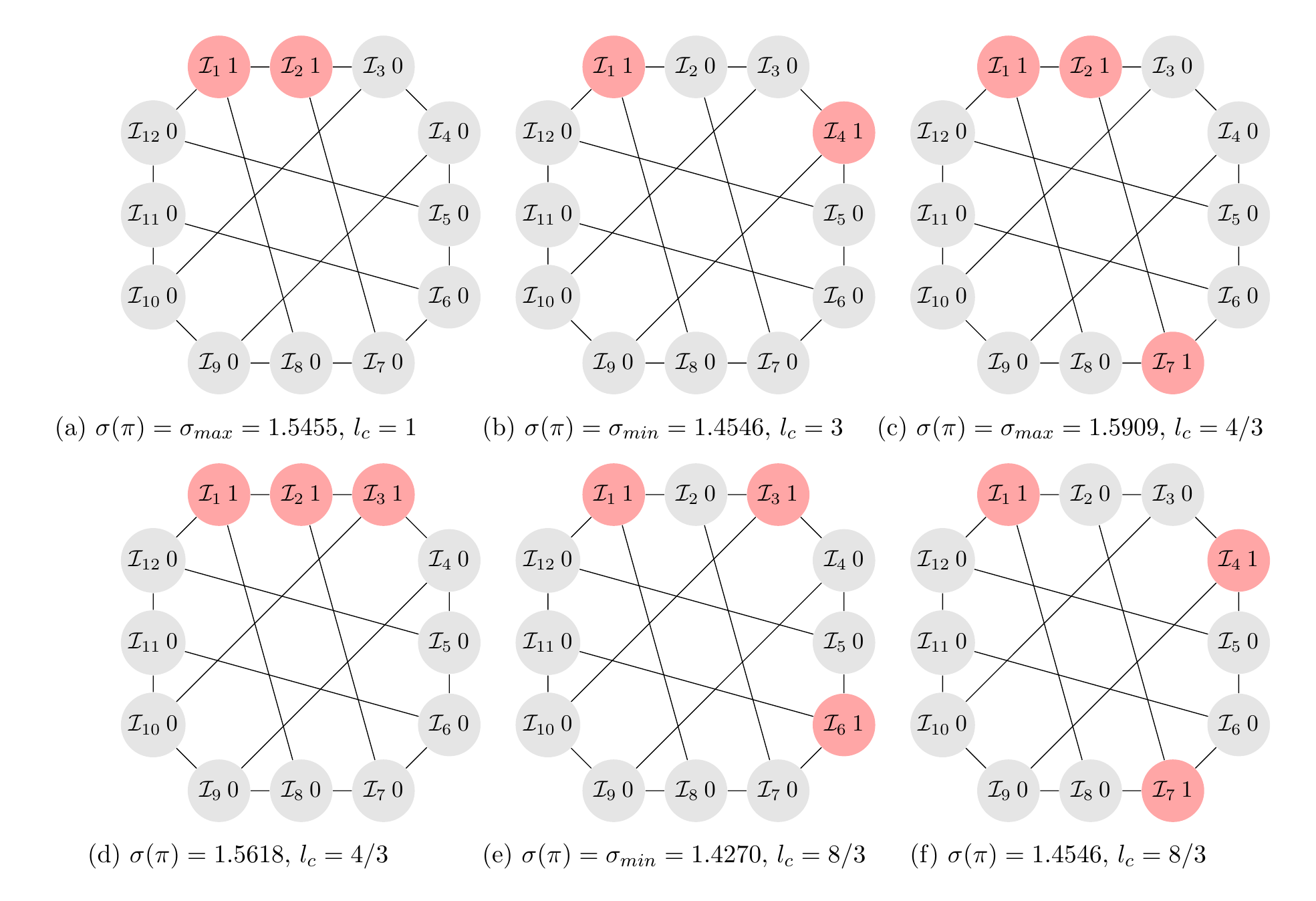}

\caption{\small{Configurations $\pi$, structure coefficients $\sigma(\pi)$ and cooperator path lengths $l_c$ for the Franklin graph~\cite{frank34}, which is vertex--transitive and triangle free. (a)--(b):  Configurations with 2 cooperators ($c(\pi)=2$).  (c)--(f): Configurations with 3 cooperators.  
}}
\label{fig:loc_frankl}
\end{figure}

3. The Franklin graph~\cite{frank34}, which is vertex--transitive and triangle free. Fig. \ref{fig:loc_frankl}(a)--(b):  Configurations with 2 cooperators ($c(\pi)=2$).  There are  $\#_{\sigma_{max}}=18$ out of 66 configurations that have the maximal value $\sigma_{max}=1.5455$, all of these (and only these) configurations have the minimal $l_c=1$, see Fig. \ref{fig:loc_frankl}(a).  Note that the maximal structure coefficient $\sigma(\pi)$ is smaller than for the Frucht and truncated tetrahedral graph (Figs. \ref{fig:loc_fre} and \ref{fig:loc_tetra}), which is $\sigma_{max}=1.5846$ obtained for configurations within a triangle of the graph. The Franklin graph is triangle free and only the second highest value of $\sigma(\pi)$ is obtained. The minimal value of $\sigma_{min}=1.4546$ is obtained for all 18 configurations with the maximal $l_c=3$,  Fig. \ref{fig:loc_frankl}(a) with $\pi=(1001\:0000\:0000)$. There are $\#_\sigma=4$ different values of $\sigma(\pi)$.   Fig. \ref{fig:loc_frankl}(c)--(f): Configurations with 3 cooperators.  The maximal value of $\sigma_{max})=1.5909$ is obtained for 12 out of 220 configurations, all have the minimal $l_c=4/3$ and additionally belong to one of the 3 squares of the Franklin graph,  $(\mathcal{I}_1,\mathcal{I}_2,\mathcal{I}_7,\mathcal{I}_8)$, $(\mathcal{I}_3,\mathcal{I}_4,\mathcal{I}_9,\mathcal{I}_{10})$, $(\mathcal{I}_5,\mathcal{I}_6,\mathcal{I}_{11},\mathcal{I}_{12})$, see Fig. \ref{fig:loc_frankl}(c). The minimal $l_c=4/3$ in itself only yields the second  largest $\sigma(\pi)=1.5618$, Fig. \ref{fig:loc_frankl}(d). The minimal value of $\sigma_{min}=1.4270$ is obtained for 24 configurations with the largest value of $l_c=8/3$ and not 2 out of 3 cooperators belonging to the same square, Fig. \ref{fig:loc_frankl}(e). However, if $l_c=8/3$ and 2 of the cooperators belong to the same square of the graph, we get the second smallest value $\sigma(\pi)=1.4546$, Fig. \ref{fig:loc_frankl}(f).
In total, there are $\#_\sigma= 8$ different values of $\sigma(\pi)$.

\begin{figure}

\includegraphics[trim = 10mm 80mm 0mm 80mm,clip,width=17.25cm, height=11.25cm]{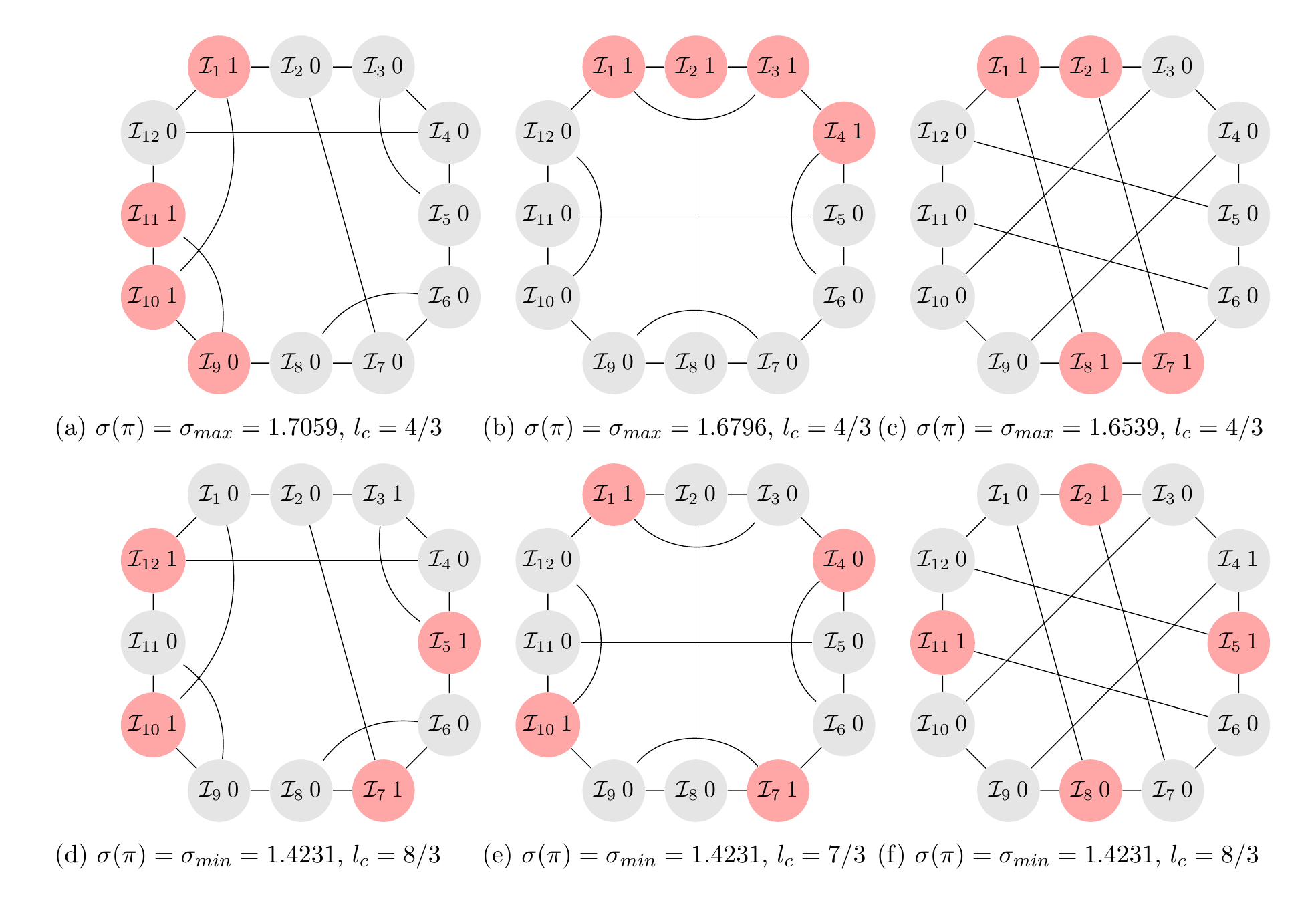}

\caption{\small{Configurations $\pi$, structure coefficients $\sigma(\pi)$ and cooperator path lengths $l_c$ for 4 cooperators ($c(\pi)=4$).
}}
\label{fig:loc_4coop}
\end{figure}

4. Configurations $\pi$, structure coefficients $\sigma(\pi)$ and cooperator path lengths $l_c$ for 4 cooperators ($c(\pi)=4$).
There are $\#_4=495$ configurations with $c(\pi)=4$.  The maximal structure coefficient $\sigma_{max}=1.7059$ for the Frucht graph is obtained for 2 configurations, both with the minimal value $l_c=4/3$ \emph{and} additionally overlapping a triangle and the square of the Frucht graph, $(\mathcal{I}_1,\mathcal{I}_{10},\mathcal{I}_{11},\mathcal{I}_{12})$, see Fig. \ref{fig:loc_4coop}(a) for the example $\pi=(1000\:0000\:1110)$, the other configuration  is $\pi=(0000\:0000\:1111)$. For the truncated tetrahedral graph, the maximal $\sigma_{max}=1.6796$ is obtained for all the 12 configurations with the minimal $l_c=4/3$, Fig. \ref{fig:loc_4coop}(b). The Franklin graph has 3 configurations with maximal $\sigma_{max}=1.6539$, they have minimal $l_c=4/3$ and each form one of the 3 squares of the graph, Fig. \ref{fig:loc_4coop}(c). The smallest values of $\sigma(\pi)$ are obtained as follows. For the Frucht graph there are 23 configurations with the smallest value $\sigma_{min}=1.4231$, all with large values of  $l_c$, (1 configurations with $l_c=8/3$, 10 configurations with $\l_c=5/2$ and 12 configurations with $l_c=7/3$), see  the example of $\pi=(0000\:1010\:0101)$ with $l_c=8/3$ in Fig. \ref{fig:loc_4coop}(d). However, there are 3 more configurations with the highest values $l_c=8/3$ that have larger values of $\sigma(\pi)$, but for these the  individual path lengths between cooperators is widely distributed, compare to  Fig.\ref{fig:loc_fre}(e) and (f). For the truncated tetrahedral graph, there are 30 configurations with the smallest value $\sigma_{min}=1.4231$ with the seconds largest value $l_c=7/3$, see the example $\pi=(1001\:0010\:0100)$ in Fig. \ref{fig:loc_4coop}(e). However, there are also 24 configurations with the largest value $l_c=5/2$ that have the second smallest value $\sigma(\pi)=1.4340$. For the Franklin graph, there are 15 configurations with the smallest value $\sigma_{min}=1.4231$, 3 of them with the largest value $l_c=8/3$ and the remaining with the second largest $l_c=5/2$, see the example $\pi(0100\:1001\:0010)$ in Fig. \ref{fig:loc_4coop}(f).


\end{document}